\documentclass[a4paper]{article}

\usepackage[T1]{fontenc}
\usepackage[utf8]{inputenc}
\usepackage[margin=1.0in]{geometry}

\usepackage{amsmath}
\usepackage{amsthm}
\usepackage{amssymb}
\usepackage{dsfont}
\usepackage{enumitem}
\usepackage{float}
\usepackage[noblocks]{authblk}
\usepackage[hidelinks]{hyperref} 
\usepackage{graphicx}
\graphicspath{{imgs/}}
\usepackage{epstopdf}
\usepackage{caption}
\usepackage[ruled,linesnumbered,noresetcount]{algorithm2e}
\usepackage{color}
\usepackage{soul}
\usepackage[natbibapa]{apacite}

\usepackage{setspace}
\SetKwRepeat{Do}{do}{while}%

\newtheorem{theorem}{Theorem}[section]
\newtheorem{lemma}[theorem]{Lemma}

\newtheorem{proposition}[theorem]{Proposition}
\theoremstyle{definition}\newtheorem{definition}[theorem]{Definition}
\theoremstyle{definition}\newtheorem{problem}{Problem}
\theoremstyle{definition}
\theoremstyle{definition}\newtheorem{assumption}[theorem]{Assumption}
\theoremstyle{definition}\newtheorem{remark}[theorem]{Remark}
\newcommand{\eq}[1]{\begin{align}#1\end{align}}
\newcommand{\eqn}[1]{\begin{align*}#1\end{align*}}
\newcommand{\demi}{\frac{1}{2}}
\newcommand{\lb}{\langle}
\newcommand{\rb}{\rangle}
\newcommand{\R}{\mathbb R}
\newcommand{\N}{\mathbb N}
\newcommand{\E}{\mathbb E}
\newcommand{\bP}{\mathbb P}
\newcommand{\bF}{\mathbb F}
\newcommand{\bS}{\mathbb S}
\newcommand{\cA}{{\bar A}}
\newcommand{\cB}{{\bar B}}
\newcommand{\cF}{\mathcal F}
\newcommand{\cP}{\mathcal P}
\newcommand{\cM}{\mathcal M}
\newcommand{\cV}{\mathcal V}
\newcommand{\cX}{\mathcal X}
\newcommand{\dt}{\partial_t}
\newcommand{\dz}{\partial_Z}
\newcommand{\dzz}{\partial_{ZZ}}
\newcommand{\dv}{\partial_V}
\newcommand{\dvv}{\partial_{VV}}
\newcommand{\dzv}{\partial_{ZV}}

\newcommand{\Dx}{\nabla_x}
\newcommand{\Dxx}{\nabla^2_x}
\newcommand{\EP}{\E^\bP}
\newcommand{\intT}{\int_0^T}
\newcommand{\intRd}{\int_{\R^d}}	
\newcommand{\xbar}{\bar{x}}

\newcommand{\etabar}{\bar{\eta}}

\newcommand{\bbeta}{\bar{\beta}}

\newcommand{\trho}{\tilde{\rho}}
\newcommand{\tcA}{{\tilde A}}
\newcommand{\tcB}{{\tilde B}}

\newcommand{\norm}[1]{\lVert #1 \rVert}
\newcommand{\iit}[1]{{\it #1}}
\newcommand{\bbf}[1]{{\bf #1}}
\newcommand{\braket}[2]{\langle#1,#2\rangle}

\begin{document}
\title{Calibration of Local-Stochastic Volatility Models by Optimal Transport}

\author[1,2]{Ivan Guo}
\author[1,2]{Gr\'egoire Loeper}
\author[1]{Shiyi Wang}
\affil[1]{School of Mathematics, Monash University, Australia}
\affil[2]{Centre for Quantitative Finance and Investment Strategies, \protect\\ Monash University, Australia}

\date{First version: June 15, 2019 \\
Revised version: \today}
\maketitle

\begin{abstract}
In this paper, we study a semi-martingale optimal transport problem and its application to the calibration of Local-Stochastic Volatility (LSV) models. Rather than considering the classical constraints on marginal distributions at initial and final time, we optimise our cost function given the prices of a finite number of European options. We formulate the problem as a convex optimisation problem, for which we provide a PDE formulation along with its dual counterpart. Then we solve numerically the dual problem, which involves a fully non-linear Hamilton--Jacobi--Bellman equation. The method is tested by calibrating a Heston-like LSV model with simulated data and foreign exchange market data. 

~
\end{abstract}

\section{Introduction}

Since the introduction of the Black--Scholes model, a lot of effort has been put on developing sophisticated volatility models that properly capture the market dynamics. In the space of equities and currencies, the most widely used models are the Local Volatility (LV) model by \citet{Dupire1994pricing} and the Stochastic Volatility (SV) models \citep[see e.g.,][]{gatheral2011volatility, heston1993closed}. Introduced as an extension of the Black--Scholes model, the LV model can be exactly calibrated to any arbitrage-free implied volatility surface. Despite this feature, the LV model has  often been criticised for its unrealistic volatility dynamics. The SV models tend to be more consistent with the market dynamics, but they struggle to fit short term market smiles and skews, and being parametric, they do not have enough degrees of freedom to match all vanilla market prices. A better fit can be obtained by increasing the number of stochastic factors in the SV models; however, this also increases the complexity of calibration and pricing.

Local-Stochastic Volatility (LSV) models, introduced in \citet{jex1999}, naturally extend and take advantage of both approaches. The idea behind LSV models is to incorporate a local, non-parametric, factor into the SV models. Thus, while keeping consistent dynamics, the model can match all observed market prices (as long as one restricts to European claims).  The determination of this local factor (also called \iit{leverage}) is based on the mimicking theorem by \citet{gyongy1986mimicking}. Research into the numerical calibration of LSV models has been developed in two different directions. One is based on a Monte Carlo approach, with \citet{henry2009calibration}, followed by \citet{guyon2011smile} using a so-called McKean's particle method. Another approach relies on solving the Fokker--Planck equation as in \citet{ren2007calibrating}. \citet{engelmann2021calibration} used the finite volume method (FVM) to solve the partial differential equation (PDE), while \citet{tian2015calibrating} considered time-dependent parameters. In a more recent study, \citet{wyns2017finite} considered a method that combines the FVM with alternating direction implicit (ADI) schemes.

All of the calibration methods mentioned above require a priori knowledge of the Local Volatility surface. This is usually obtained by using Dupire's formula \citep{Dupire1994pricing} assuming the knowledge of vanilla options for all strikes and maturities. However, only a finite number of options are available in practice. Thus, an interpolation of the implied volatility surface or option prices is often needed, which can lead to inaccuracies and instabilities. Inaccuracies can come from the usage of a parametric model for the volatility surface that will not match perfectly market prices by definition. Instabilities can come form the interpolating model being not arbitrage-free. It also raises the question of what arbitrary shape of extrapolation one is going to take for very out of the money strikes. Moreover, there is no a priori control on the regularity of the leverage function, and even its very existence remains an open problem, although some results for small times have been obtained in \citet{AbergelTachet} \citep[cf.][for an application of Tikhonov regularisation technique to the LSV calibration problem]{saporito2017calibration}. Other related works include \citet{lacker2019inverting, jourdain2020existence}. In a recent work of \citet{cuchiero2020generative}, the LSV calibration problem was addressed from a deep learning point of view. In particular, the leverage function is parameterised by a class of feed-forward neural networks, and the model is calibrated by a generative adversarial network approach. In the present work, inspired by the theory of optimal transport, we introduce a variational approach for calibrating LSV models that does not require any form of interpolation. 

In recent years, optimal transport theory has attracted the attention of many researchers. The problem was first addressed by \citet{monge1781memoire} in the context of civil engineering and was later given a modern mathematical treatment by \citet{kantorovich1948}. In 2000, in a landmark paper, \citet{benamou-brenier2000} introduced a time-continuous formulation of the problem, which they solved numerically by an augmented Lagrangian method. In \citet{brenier1999} and \citet{loeper2006}, the dual formulation of the time-continuous optimal transport has been formally expressed and generalised as an application of the Fenchel--Rockafellar theorem \citep[see e.g.,][Theorem 1.9]{villani2003book}. This method has been also applied in \citet{huesmann2017} to study a time-continuous formulation of the martingale optimal transport. Recently, the problem has been extended to transport by semi-martingales. \citet{tan-touzi2013} studied the optimal transport problem for semi-martingales with constraints on the marginals at initial and final times. More recently, in \citet{guo2018path}, the semi-martingale optimal transport problem was further extended to a more general path-dependent setting. 

In the area of mathematical finance, optimal transport theory has recently been applied to many different problems, see e.g., \citet{henry2016explicit,dolinsky-soner2014, pal2018exponentially}. In terms of volatility model calibration by optimal transport, in the previous work \citep{guo2017local}, the authors explored the idea of calibrating LV models to European options, in which they adapted the augmented Lagrangian method of \citet{benamou-brenier2000} to one-dimensional martingale optimal transport. Later in \citet{guo2018path}, the first two authors expanded the calibrating instruments from European options to path-dependent options, such as Asian options, barrier options and lookback options. We also mention that a variational calibration method of the LV model was proposed in \citet{avellaneda1997calibrating} much earlier, although the connection with optimal transport was not established at that time. In addition to the LV models, a new class of SV models were developed, in \citet{henry2019martingale}, inspired by the so-called Schr\"odinger bridge problem which is closely related to optimal transport. The calibration of these new SV models is achieved by modifying only the drift and leaving the volatility of volatility unchanged. Apart from continuous models, as an extension of \citet{march2019building}, \citet{guyon2020joint} constructed a discrete-time model which solves the challenging joint calibration problem of SPX and VIX options. Recently in \citet{guo2020joint}, we propose a continuous-time model to solve also the joint calibration problem by extending the approach of this paper.

In this paper, we further extend the approach of \citet{guo2017local} and \citet{guo2018path} to the calibration of LSV models. The calibration problem is formulated as a semi-martingale optimal transport problem. Unlike \citet{tan-touzi2013}, we consider a finite number of discrete constraints given by the prices of European claims. As a consequence of Jensen's inequality, we show that an optimal diffusion process can be chosen to be Markovian in the state variables given by the initial SV model. This result leads to a PDE formulation. By following the duality theory of optimal transport introduced in \citet{brenier1999} and a smoothing argument used in \citet{bouchard2017impact}, we establish a dual formulation. We also provide a numerical method to solve a fully non-linear Hamilton--Jacobi--Bellman (HJB) equation arising in the dual formulation. Finally, numerical examples show that the model can be fully calibrated to the European options with both simulated data and FX market data.

Despite its accuracy, our method is quite demanding in terms of computational power. The gradient descent demands at each step to solve one non-linear 2$d$ PDE, and the computation of the gradient requires one (linear) 2$d$ PDE per instrument. The most costly operation of numerically solving a linear PDE is inverting a large sparse matrix. However, this operation only needs to be carried once per time step because the computations of all components of the gradient are computed by solving the same linear PDE but with different terminal conditions. Alternatively, all gradients can be efficiently computed in one Monte Carlo simulation, which is a choice we did not make here for the sake of accuracy. Going into higher dimensions (a multi-factor stochastic volatility model for example) would require to increase the dimension of the PDEs, which is problematic as soon as $d\geq 3$. When the goal is only to solve the usual LSV calibration problem, i.e., to find the leverage function, other methods achieve the result faster. For example, for a one-factor model, the PDE method of \citet{ren2007calibrating} only requires solving a two-dimensional non-linear PDE once, and the particle method of \citet{guyon2011smile} is even faster and can be applied to high dimensional cases (e.g., calibrating an LSV model with multiple stochastic factors). On the other hand, with the technique developed in this paper, one can fit path-dependent products \citep{guo2018path}, SPX and VIX options \citep{guo2020joint} and here LSV models. Therefore the interest of our method is clearly its broad range of applications, at the cost of a relatively heavy computational cost. We also believe that with the recent developments of numerical methods for solving non-linear PDEs in high dimensions \citep[see e.g.,][]{han2020algorithms}, our method can be greatly improved in terms of computational speed, and become applicable in high dimensions. Also notice that, being based on gradient descent, for a slight update of the market data, only a few gradient iterations should be needed to update the model. Finally, our method provides a rigorous existence result of an LSV type model. Previous works by \citet{AbergelTachet} only provide an existence result for small times (see also \citet{lacker2019inverting, jourdain2020existence}).

The paper is organised as follows: In Section 2, we introduce some preliminary definitions. In Section 3, we show the connection between the semi-martingale optimal transport problem and a PDE formulation. Duality results are then established for the PDE formulation. In Section 4, we demonstrate the calibration method using a Heston-like LSV model. Numerical method and results with both simulated data and FX market data are provided in Section 5.

\section{Preliminaries}

Given a Polish space $E$ equipped with its Borel $\sigma$-algebra, let $C(E)$ be the space of continuous functions on $E$ and $C_b(E)$ be the space of bounded continuous functions. Denote by $\cM(E)$ the space of finite signed Borel measures endowed with the weak-$*$ topology. Let $\cM_+(E)\subset\cM(E)$ denote the subset of nonnegative measures. If $E$ is compact, the topological dual of $C_b(E)$ is given by $C_b(E)^*=\cM(E)$. More generally, if $E$ is non-compact, $C_b(E)^*$ is larger than $\cM(E)$. Let $\cP(E)$ be the space of Borel probability measures, $BV(E)$ be the space of functions of bounded variation and $L^1(d\mu)$ be the space of $\mu$-integrable functions. We also write $C_b(E,\R^d)$, $\cM(E,\R^d)$, $BV(E,\R^d)$ and $L^1(d\mu,\R^d)$ as the vector-valued versions of their corresponding spaces. If $\mu_t(x)=\mu(t,x)$ is a measure defined on $[0,T]\times\R^d$, we will write $d\mu$ or $d\mu_tdt$ in short for $\mu(t,dx)dt$. Denote by $\bS^d$ the set of $d\times d$ symmetric matrices and $\bS_+^d\subset \bS^d$ the set of positive semidefinite matrices. For any matrices $A,B\in\bS^d$, we write $A:B:=\operatorname{tr}(A^\intercal B)$ for their scalar product. For convenience, let $\Lambda=[0,T]\times\R^d$ and $\cX=\R\times\R^d\times\bS^d$. We use the notation $\braket{\cdot}{\cdot}$ to denote the duality bracket between $C_b(\Lambda,\cX)$ and $C_b(\Lambda,\cX)^*$.

Let $\Omega:=C([0,T],\R^d), T>0$ be the canonical space with the canonical process $X$ and the canonical filtration $\bF=(\cF_t)_{0\leq t\leq T}$ generated by $X$. We denote by $\cP$ the collection of all probability measures $\bP$ on $(\Omega, \cF_T)$ under which $X\in\Omega$ is an $(\bF,\bP)$-semi-martingale given by
\eqn{
  X_t = X_0 + A_t + M_t, \quad t\in [0,T], \quad \bP\text{-a.s.,}
}
where $M$ is an $(\bF,\bP)$-martingale with quadratic variation $\lb X_t \rb = \lb M_t \rb = B_t$, and the processes $A$ and $B$ are $\bP$-a.s. absolutely continuous with respect to $t$. We say $\bP$ is \iit{characterised} by $(\alpha^\bP, \beta^\bP)$ if 
\eqn{
  \alpha_t^\bP=\frac{dA_t^\bP}{dt},\quad \beta_t^\bP=\frac{dB_t^\bP}{dt},
}
where $(\alpha^\bP, \beta^\bP)$ take values in the space $\R^d\times\bS^d_+$. Note that $(\alpha^\bP, \beta^\bP)$ is $\bF$-adapted and determined up to $d\bP\times dt$, almost everywhere. Let $\cP^1\subset\cP$ be the subset of probability measures $\bP$ under which the characteristics $(\alpha^\bP, \beta^\bP)$ are $\bP$-integrable on the interval $[0,T]$. In other words,
\eqn{
  \E^\bP\left(\intT|\alpha_t^\bP|+|\beta_t^\bP|\,dt\right)<+\infty,
}
where $|\cdot|$ is the $L^1$-norm.

Given a vector $\tau:=(\tau_1,\ldots,\tau_m)\in(0,T]^m$, denote by $G$ a vector of $m$ functions such that each function $G_i\in C_b(\R^d)$ for $i=1,\ldots,m$. Given a Dirac measure $\mu_0=\delta_{x_0}$ and a vector $c\in\R^m$, we define $\cP(\mu_0,\tau,c,G)\subset \cP^1$ as follows:
\eqn{
  \cP(\mu_0,\tau,c,G):=\{ \bP:\bP\in\cP^1,\, \bP\circ X^{-1}_0 = \mu_0 \text{ and } \EP[G_i(X_{\tau_i})]=c_i,\, i=1,\ldots,m \}.
}

\begin{assumption}
  The final time $T$ coincides with the longest maturity, i.e., $T=\max_k \tau_k$.
\end{assumption}

For technical reasons, we restrict ourselves to functions $G_i$ in $C_b(\R^d)$.  In the context of volatility models calibration, $G_i$ are discounted European payoffs. Although the call option payoff functions are not technically in $C_b(\R^d)$, we only work with them in a truncated (compact) space in practice. Alternatively, one may consider only put options using put-call parity. It is possible to relax the assumption $G_i\in C_b(\R^d)$, but it would require a different set up in topological spaces.

\section{Main results}\label{sec:3}
\subsection{Formulations}
In this section, we first formulate the semi-martingale optimal transport problem under discrete constraints. Then a PDE formulation is introduced along with its dual counterpart. 

Define the cost function $F:\Lambda\times\R^d\times\bS^d \to \R\cup\{+\infty\}$ where $F(t,x,\alpha,\beta)=+\infty$ if $\beta\notin\bS^d_+$, and $F(t,x,\alpha,\beta)$ is nonnegative, proper, lower semi-continuous, strongly convex and coercive in $(\alpha,\beta)$ and uniformly in $(t,x)$. By $F$ being strongly convex in $(\alpha,\beta)$ we mean that there exists a constant $C>0$ such that for all $t,x,\alpha,\beta,\alpha',\beta'$ and any subderivative $\nabla F$, where $\nabla$ is performed over $(\alpha,\beta)$, if $F(t,x,\alpha,\beta)$ is finite then
\eqn{
  F(t,x,\alpha',\beta')\geq F(t,x,\alpha,\beta) + \braket{\nabla F(t,x,\alpha,\beta)}{(\alpha'-\alpha, \beta'-\beta)} + C(\norm{\alpha'-\alpha}^2 + \norm{\beta'-\beta}^2),
}
where $\norm{\cdot}$ denotes the Euclidean norm on $\R^d$ and $\bS^d$. By $F$ being coercive in $(\alpha,\beta)$ we mean that there exist constants $p>1$ and $C>0$ for all $t,x,\alpha,\beta$ such that
\eqn{
  |\alpha|^p + |\beta|^p \leq C(1 + F(t,x,\alpha,\beta)).
}
The convex conjugate of $F$ with respect to $(\alpha,\beta)$ is denoted by $F^*:\Lambda\times\R^d\times\bS^d \to \R\cup\{+\infty\}$ and is given by 
\eq{\label{eq:F_conjugate}
  F^*(t,x,a,b) := \sup_{\alpha\in\R^d, \beta\in \bS^d} \left\{\alpha\cdot a + \beta:b - F(t,x,\alpha,\beta)\right\}.
}
We remark that $\bS^d$ in \eqref{eq:F_conjugate} can be replaced by $\bS^d_+$ due to the assumption that $F(t,x,\alpha,\beta)$ is finite only if $\beta\in\bS^d_+$. For simplicity, we write $F(\alpha,\beta):=F(t,x,\alpha,\beta)$ and $F^*(a,b):=F^*(t,x,a,b)$ if there is no ambiguity. Note that our definition of strongly convex does not require $F$ to be differentiable, since only subderivatives are used. Nevertheless, it implies that $F$ is strictly convex and thus $F^*$ is differentiable. In addition, the coercivity of $F$ implies that $F^*$ is finite.

Adopting the convention $\inf\emptyset=+\infty$, we are interested in the following minimisation problem:
\begin{problem}\label{prob:main}
Given $\mu_0, \tau, c$ and $G$, we want to find
\eqn{
  \cV = \inf_{\bP\in\cP(\mu_0,\tau,c,G)}\EP \int_0^T F(\alpha_t^\bP,\beta_t^\bP) \,dt.
}
The problem is said to be \iit{admissible} if $\cP(\mu_0,\tau,c,G)$ is nonempty and the infimum above is finite.
\end{problem}

It is well known that the marginal distributions of diffusion processes at fixed times solve the Fokker--Planck equation in the weak sense. The converse result was given by \citet{Figalli2008existence} and \citet{Trevisan2016existence}. For brevity, we write $\EP_{t,x}:=\EP[\,\cdot\mid X_t=x]$. As an immediate consequence of It\^o's formula and Theorem 2.5 in \citet{Trevisan2016existence}, we introduce the following lemma.
\begin{lemma}\label{lemma:3.2}
  Let $\bP\in\cP^1$ and $\rho^\bP_t=\rho^\bP(t,\cdot) = \bP\circ X_t^{-1}$ be the marginal distribution of $X_t$ under $\bP$, $t\leq T$. Then $\rho^\bP$ is a weak solution to the Fokker--Planck equation:
  \eq{\label{eq:fokker-planck}
  \left\{\begin{array}{r@{\ }c@{\ }l@{\ }l} \displaystyle
    \dt\rho^\bP_t + \Dx\cdot(\rho^\bP_t\E^\bP_{t,x}\alpha^\bP_t) - \demi\sum_{i,j}\partial_{i j}(\rho^\bP_t(\E^\bP_{t,x}\beta^\bP_t)_{ij}) &=& 0  & \quad\mbox{in }[0, T]\times\R^d, \\
    \rho^\bP_0 &=& \delta_{X_0} & \quad\mbox{in }\R^d.
  \end{array}\right.
  }
  Moreover, there exists another probability measure $\bP'\in\cP^1$, characterised by $(\alpha^{\bP'}, \beta^{\bP'})$, under which $X$ has the same marginals, $\rho^{\bP'}=\rho^{\bP}$, and is a Markov process solving 
  \eq{\label{eq:localised-sde}
  \left\{\begin{array}{r@{\ }c@{\ }l} \displaystyle
   dX_t &=& \alpha^{\bP'}(t,X_t)dt + (\beta^{\bP'}(t,X_t))^\demi\,dW_t^{\bP'},\quad 0\leq t\leq T, \\
   X_0 &=& x_0,
  \end{array}\right.
  }
  where $W^{\bP'}$ is a $\bP'$-Brownian motion, $\alpha^{\bP'}(t,X_t)=\E^\bP_{t,X_t}\alpha^\bP_t$ and $\beta^{\bP'}(t,X_t)=\E^\bP_{t,X_t}\beta^\bP_t$.
\end{lemma}
The above lemma provides a solution to study semi-martingales via Markov processes in the form of \eqref{eq:localised-sde}. It is worth noting that the idea of using diffusion processes to mimic an It\^o process by matching their marginals at fixed times traces back to the classical mimicking theorem of \citet{gyongy1986mimicking}. The uniform ellipticity condition of Gy\"ongy's mimicking theorem was later relaxed by \citet{Brunick2013mimicking}. In fact, if $X$ is an It\^o process under $\bP$, Lemma \ref{lemma:3.2} can be seen as a reformulation of \citet[Corollary 3.7]{Brunick2013mimicking} which was constructed by a completely different approach. The Markov processes $X$ in \eqref{eq:localised-sde} are also called Markovian projections in the literature. Note that, in \citet{Brunick2013mimicking}, even though the main results are given for It\^o processes, the authors first provide more general results for semi-martingales (see \citet[Theorem 7.1]{Brunick2013mimicking}) and then prove the main results for It\^o processes by the It\^o representation theorem. Therefore, Lemma \ref{lemma:3.2} can also be proved by the results of \citet{Brunick2013mimicking}.

\begin{definition}
  Define $\cP_{loc}(\mu_0,\tau,c,G)$ to be the subset of $\cP(\mu_0,\tau,c,G)$ such that, under any $\bP\in\cP_{loc}(\mu_0,\tau,c,G)$, $X$ is a Markov process that takes the form of (\ref{eq:localised-sde}).
\end{definition}

\begin{lemma}\label{lemma:3.4}
  If $\cP(\mu_0,\tau,c,G)$ is not empty, then $\cP_{loc}(\mu_0,\tau,c,G)$ is not empty. Moreover, for any $\bP\in\cP(\mu_0,\tau,c,G)$, there exists a $\bP'\in\cP_{loc}(\mu_0,\tau,c,G)$ such that $X$ has the same marginals under $\bP$ and $\bP'$.
\end{lemma}
\begin{proof}
  Assume that $\cP(\mu_0,\tau,c,G)$ is not empty, for any $\bP\in\cP(\mu_0,\tau,c,G)$, by Lemma \ref{lemma:3.2}, there exists $\bP'\in\cP^1$ such that $X$ is a Markov process that has the same marginals $\rho^{\bP'}=\rho^\bP$ and takes the form of (\ref{eq:localised-sde}) with coefficients $(\alpha^{\bP'}(t,X_t),\beta^{\bP'}(t,X_t))=(\EP_{t,X_t}\alpha_t^\bP,\EP_{t,X_t}\beta_t^\bP)$. Since $\rho^{\bP'}=\rho^\bP$, $X$ has the initial marginal $\mu_0$ and satisfies $\E^{\bP'}[G_i(X_{\tau_i})]=c_i$ for all $i=1,\ldots,m$ under both $\bP$ and $\bP'$. Thus, $\bP'\in\cP_{loc}(\mu_0,\tau,c,G)$.
\end{proof}

Applying Lemma \ref{lemma:3.4} and taking advantage of the convexity of the cost function, we establish the following result:
\begin{proposition}\label{prop:3.3}
Given $\mu_0, \tau, c$ and $G$, then
\eq{\label{eq:3}
  \cV = \inf_{\bP\in\cP(\mu_0,\tau,c,G)}\EP \intT F(\alpha_t^\bP,\beta_t^\bP) \,dt = \inf_{\bP\in\cP_{loc}(\mu_0,\tau,c,G)}\E^{\bP}\intT F(\alpha^\bP(t,X_t),\beta^\bP(t,X_t)) \, dt.
}
\end{proposition}
\begin{proof}
  If $\cP(\mu_0,\tau,c,G)$ is empty, then $\cP_{loc}(\mu_0,\tau,c,G)$ is empty since $\cP_{loc}(\mu_0,\tau,c,G)\subset\cP(\mu_0,\tau,c,G)$. Thus, \eqref{eq:3} holds and $\cV = +\infty$. 
  
  If $\cP(\mu_0,\tau,c,G)$ is not empty, by Lemma \ref{lemma:3.4}, $\cP_{loc}(\mu_0,\tau,c,G)$ is not empty. For any $\bP\in\cP(\mu_0,\tau,c,G)$, let $\bP'\in\cP_{loc}(\mu_0,\tau,c,G)$ be a probability measure such that $X$ has the same marginals under $\bP$ and $\bP'$. Applying Jensen's inequality together with the tower property of conditional expectation, we have
  \begin{equation}
  \begin{aligned}\label{eq:local_ineq}
    \EP\intT F(\alpha_t^\bP,\beta_t^\bP)\,dt &= \EP\intT \EP_{t,X_t}F(\alpha_t^\bP,\beta_t^\bP) \,dt \\
          &\geq \EP\intT F(\EP_{t,X_t}\alpha_t^\bP,\EP_{t,X_t}\beta_t^\bP)\, dt \\
          &= \E^{\bP'}\intT F(\alpha^{\bP'}(t,X_t),\beta^{\bP'}(t,X_t)) \, dt.
  \end{aligned}
  \end{equation}
  The last $\E^{\bP}$ is replaced by $\E^{\bP'}$ because the marginal of $X$ is the same under $\bP$ and $\bP'$. Since $\cP_{loc}(\mu_0,\tau,c,G)\subset\cP(\mu_0,\tau,c,G)$, taking infimum over all $\bP\in\cP(\mu_0,\tau,c,G)$ on the left-hand side and over all $\bP'\in\cP_{loc}(\mu_0,\tau,c,G)$ on the right-hand side of (\ref{eq:local_ineq}), we obtain the required result.
\end{proof}

Proposition \ref{prop:3.3} shows that it suffices to consider only the probability measures in $\cP_{loc}(\mu_0,\tau,c,G)$. Thus, by the connections established in Lemma \ref{lemma:3.2}, Problem 1 can be studied via PDE methods. Following the Benamou--Brenier formulation of the classical optimal transport from \citet{benamou-brenier2000}, we introduce the following problem:
\begin{problem}[PDE formulation]\label{prob:pde}
Given $\mu_0, \tau, c$ and $G$, we want to solve
\eq{\label{eq:p2_obj}
  \cV=\inf_{\rho,\alpha,\beta}\intT\intRd F(\alpha(t,x),\beta(t,x))\,\rho(t,dx)dt,
}
among all $(\rho,\alpha,\beta)\in C([0,T],\cP(\R^d)-w*)\times L^1(d\rho_tdt,\R^d)\times L^1(d\rho_tdt,\bS^d)$ satisfying (in the distributional sense)
\eq{
  \dt\rho(t,x) + \Dx\cdot(\rho(t,x)\alpha(t,x)) - \demi\sum_{i,j}\partial_{i j}(\rho(t,x)\beta_{ij}(t,x)) = 0, \label{eq:p2_cons_1}\\
  \intRd G_i(x)\,\rho(\tau_i,dx) = c_i,\; \forall i=1,\ldots,m, \qquad\text{and}\qquad
  \rho(0,\cdot)= \mu_0. \label{eq:p2_cons_2}
}
\end{problem}
The interchange of integrals in (\ref{eq:p2_obj}) is justified by Fubini's theorem as $F$ is nonnegative. For the weak continuity of measure $\rho$ in time, the reader can refer to \citet[Theorem 3]{loeper2006}.

Based on the results of Sections \ref{sec:duality} and \ref{sec:viscosity} below, we shall introduce a dual formulation of Problem \ref{prob:pde}. In the proposition below, $C_b^2(\R^d)$ is the space of twice continuously differentiable functions with bounded partial derivatives up to order 2, and it is equipped with the norm given by the supremum of all partial derivatives up to order 2. The subscript of $\phi_\lambda$ indicates the implicit dependence of $\phi$ on $\lambda$ via the HJB equation. The definition of the viscosity solution to \eqref{eq:pde_dual} and the proof will be given in Section \ref{sec:viscosity}.
\begin{proposition}[Dual formulation]\label{prop:optimal_sol2}
  If Problem \ref{prob:main} is admissible, then
  \eq{\label{eq:dual_obj}
    \cV=\sup_{\lambda\in\R^m}\left\{ \sum_{i=1}^m \lambda_i c_i - \int_{\R^d} \phi_\lambda(0,x)\,d\mu_0 \right\},
  }
  where $\phi$ is the viscosity solution to the HJB equation 
  \eq{\label{eq:pde_dual}
  \dt\phi_\lambda+\sum_{i=1}^m \lambda_i G_i\delta_{\tau_i} + F^*(\Dx\phi_\lambda,\demi\Dxx\phi_\lambda) = 0, \quad \mbox{in } [0,T)\times\R^d,
  }
  with the terminal condition $\phi_\lambda(T,\cdot)=0$. Moreover, if there exists $(\rho,\alpha,\beta)\in C([0,T],\cP(\R^d)-w*)\times L^1(d\rho_tdt,\R^d)\times L^1(d\rho_tdt,\bS^d)$ satisfying \eqref{eq:p2_cons_1} and \eqref{eq:p2_cons_2} (in the distributional sense), then the infimum of Problem \ref{prob:pde} is attained. If the supremum is attained by some $\lambda^*\in\R^m$ and $(\rho,\alpha,\beta)$ is an optimal solution of Problem \ref{prob:pde}, then $(\alpha,\beta)$ is given by
  \eq{\label{eq:optimal_alpha_beta}
    (\alpha,\beta) = \nabla F^*(\Dx\phi_{\lambda^*},\demi\Dxx\phi_{\lambda^*}), \quad d\rho_tdt-\mbox{almost everywhere}.
  }
\end{proposition}
Before ending this section, it is worth commenting on the admissibility of Problem \ref{prob:main}. We have chosen to impose the admissibility assumption in order to simplify our presentation and arguments. With some modifications, it is possible to remove this assumption from the primal problem and still obtain duality. In particular, both sides of the duality would be infinite if the problem is not admissible. Then, characterising the admissibility of Problem \ref{prob:main} corresponds to checking the finiteness of the dual problem, and can be seen as a more elaborate analogue of Strassen's theorem for the classical optimal transport problem.

\subsection{Duality}\label{sec:duality}

This section is devoted to establishing the duality by closely following \citet[Section 3.2]{loeper2006} \citep[see also][]{brenier1999, huesmann2017}. 
\begin{theorem}\label{thm:dual}
If Problem \ref{prob:main} is admissible, then
\eq{\label{eq:dual_obj_thm}
  \cV=\sup_{\phi,\lambda} \left\{ \sum_{i=1}^m \lambda_i c_i - \int_{\R^d} \phi(0,x)\,d\mu_0 \right\},
}
where the supremum is taken over all $(\phi,\lambda)\in BV([0,T],C_b^2(\R^d))\times \R^m$ satisfying
\eq{\label{ieq:hjb}
  \dt\phi+\sum_{i=1}^m \lambda_i G_i\delta_{\tau_i} + F^*(\Dx\phi,\demi\Dxx\phi)&\leq 0 \qquad \mbox{in } [0,T)\times \R^d,
}
and $\phi(T,\cdot)=0$. Moreover, if there exists $(\rho,\alpha,\beta)\in C([0,T],\cP(\R^d)-w*)\times L^1(d\rho_tdt,\R^d)\times L^1(d\rho_tdt,\bS^d)$ satisfying \eqref{eq:p2_cons_1} and \eqref{eq:p2_cons_2} (in the distributional sense), then the infimum of Problem \ref{prob:pde} is attained.
\end{theorem}
\begin{proof}
The proof relies on the Fenchel--Rockafellar theorem which plays a key role in the applications of convex analysis. One may note that the objective function \eqref{eq:p2_obj} is not convex in $(\rho, \alpha,\beta)$ since $F(\alpha,\beta)\rho$ is not convex in $(\rho, \alpha,\beta)$. As we will see below, \eqref{eq:p2_obj} can be written as the convex conjugate (which is always convex) of another function with respect to $(\rho,\cA:=\alpha\rho,\cB:=\beta\rho)$ and $(\cA, \cB)$ are absolutely continuous with respect to $\rho$. In addition, the constraints \eqref{eq:p2_cons_1} and \eqref{eq:p2_cons_2} are linear in $(\rho,\cA,\cB)$. Therefore, throughout the proof, we will work on $(\rho,\cA,\cB)$ instead. For simplicity, we will write $d\cA$ and $d\cB$ in short for $\alpha(t,x)\rho(t,dx)dt$ and $\beta(t,x)\rho(t,dx)dt$, respectively.

Formulate the constraints \eqref{eq:p2_cons_1} and \eqref{eq:p2_cons_2} in the following weak form:
\eq{
  \forall\phi &\in C_c^\infty(\Lambda), & \begin{split}\int_\Lambda \dt\phi\,d\rho + \Dx\phi\cdot d\cA + \demi\Dxx\phi:d\cB + \int_{\R^d} \phi(0,\cdot)\,d\mu_0 &= 0, \\ \phi(T,\cdot)&=0\end{split}
  \label{eq:weak_cons_1} \\ 
  \forall \lambda &\in \R^m, & \int_\Lambda \sum_{i=1}^m \lambda_i G_i\delta_{\tau_i} d\rho - \sum_{i=1}^m \lambda_i c_i &= 0 .\label{eq:weak_cons_2}
}
where $C^\infty_c(\Lambda)$ is the space of smooth functions with compact support on $\Lambda$. Thus Problem 2 can be reformulated as the following saddle point problem:
\eq{\label{eq:cV_infsup}
  \begin{split}
  \cV=\inf_{\rho,\cA,\cB}\sup_{\phi,\lambda}\bigg\{ &\int_\Lambda F\left(\frac{d\cA}{d\rho},\frac{d\cB}{d\rho}\right)\,d\rho -\dt\phi\,d\rho - \Dx\phi\cdot d\cA - \demi\Dxx\phi:d\cB - \int_{\R^d} \phi(0,\cdot)\,d\mu_0 \\
  &- \int_\Lambda\sum_{i=1}^m \lambda_i G_i\delta_{\tau_i}\,d\rho + \sum_{i=1}^m \lambda_i c_i  \bigg\}.
  \end{split}
}
The strategy of the proof is to first construct a function $\Phi$ whose convex conjugate $\Phi^*$ is equal to the objective function of Problem \ref{prob:pde}, and construct another function $\Psi$ whose convex conjugate $\Psi^*$ is equal to the rest part inside the infimum of \eqref{eq:cV_infsup} so that $\cV=\inf_{\rho,\cA,\cB}(\Phi^*+\Psi^*)(\rho,\cA,\cB)$. Then, the duality is established by applying the Fenchel--Rockafellar theorem.

Adopting the terminology of \citet{huesmann2017}, we say the triple $(r,a,b)$ is \iit{represented} by $(\phi,\lambda)$ if it satisfies
\eqn{
  r+\dt\phi+\sum_{i=1}^m\lambda_i G_i\delta_{\tau_i} &= 0, \\
  a+\Dx\phi &= 0, \\
  b+\demi\Dxx\phi &= 0.
}
If we choose $(r,a,b)$ from $C_b(\Lambda,\cX)$, by the first equation above, $\dt\phi$ is a measure because of the presence of the Dirac delta functions. Thus, $\phi$ has bounded variation with respect to $t$ on $[0, T]$ and has possible jump discontinuities at $t=\tau_i$. Now, define functionals $\Phi:C_b(\Lambda,\cX)\to\R\cup\{+\infty\}$ and $\Psi:C_b(\Lambda,\cX)\to\R\cup\{+\infty\}$ as follows:
\eqn{
  \Phi(r,a,b) &= \left\{ \begin{array}{ll} 0 & \mbox{if } r+F^*(a,b)\leq 0, \\
		 +\infty & \mbox{otherwise,} \end{array} \right. \\
  \Psi(r,a,b) &= \left\{ \begin{array}{ll} \displaystyle  \int_{\R^d} \phi(0,x)\,d\mu_0 - \sum_{i=1}^m \lambda_i c_i
					&\begin{array}{l}\mbox{if $(r,a,b)$ is represented by $(\phi,\lambda)$} \\ \mbox{in $BV([0,T],C_b^2(\R^d))\times\R^m$ with $\phi(T,\cdot)=0$,} \end{array} \\
			 +\infty & \begin{array}{l} \mbox{otherwise.} \end{array} \end{array} \right.
}
Note that $\Psi$ is well-defined. If $\Psi(r,a,b)<+\infty$ for some $(r,a,b)$ that is represented by some $(\phi, \lambda)$, then $(\phi, \lambda)$ satisfies the constraints \eqref{eq:weak_cons_1} and \eqref{eq:weak_cons_2}, otherwise we can arbitrarily scale $(\phi,\lambda)$ in \eqref{eq:cV_infsup} then $\cV$ becomes unbounded. Assume that $(r,a,b)$ can be represented by both $(\hat\phi,\hat\lambda)$ and $(\tilde\phi,\tilde\lambda)$, then we have 
\eq{
  \dt(\hat\phi-\tilde\phi) + \sum_{i=1}^m(\hat\lambda_i-\tilde\lambda_i)G_i\delta_{\tau_i} = 0. \label{eq:welldefined}
}
Integrating \eqref{eq:welldefined} with any $\rho$ that satisfies \eqref{eq:weak_cons_2} and $\rho(0,\cdot)=\mu_0$, we have $\int_{\R^d} \hat\phi(0,x)\,d\mu_0 - \sum_{i=1}^m \hat\lambda_i c_i = \int_{\R^d} \tilde\phi(0,x)\,d\mu_0 - \sum_{i=1}^m \tilde\lambda_i c_i$, so the value of $\Psi$ does not depend on the choice of $(\phi, \lambda)$ and hence $\Psi$ is well-defined.

Denote by $\Phi^*$ and $\Psi^*$ the convex conjugates of $\Phi$ and $\Psi$, respectively. For $\Phi$, its convex conjugate $\Phi^*:C_b(\Lambda,\cX)^*\to\R\cup\{+\infty\}$ is given by
\eqn{
  \Phi^*(\rho,\cA,\cB) = \sup_{(r,a,b)\in C_b(\Lambda,\cX)} \{ \braket{(r,a,b)}{(\rho,\cA,\cB)} \;;\; r+F^*(a,b)\leq 0 \}.
}
As shown in Lemma \ref{lemma:A1}, if we restrict $\Phi^*$ to $\cM(\Lambda,\cX)$, then
\eqn{
  \Phi^*(\rho,\cA,\cB) &= \left\{ \begin{array}{ll} \displaystyle\int_\Lambda F\left(\frac{d\cA}{d\rho},\frac{d\cB}{d\rho}\right)\,d\rho & \mbox{if }\rho\in\cM_+(\Lambda,\cX) \mbox{ and } (\cA,\cB)\ll\rho, \\
			 +\infty & \mbox{otherwise.} \end{array} \right.
}
Next, $\Psi^*:C_b(\Lambda,\cX)^*\to\R\cup\{+\infty\}$ is given by
\eqn{
  \Psi^*(\rho,\cA,\cB) = \sup_{(r,a,b)}\left\{ \braket{(r,a,b)}{(\rho,\cA,\cB)} - \int_{\R^d} \phi(0,x)\,d\mu_0 + \sum_{i=1}^m \lambda_i c_i \right\},
}
where the supremum is taken over all triples $(r,a,b)\in C_b(\Lambda,\cX)$ represented by $(\phi,\lambda)$ in $BV([0,T],C_b^2(\R^d))\times\R^m$. In terms of $(\phi, \lambda)$, 
\eqn{
  \Psi^*(\rho,\cA,\cB) = \sup_{\phi, \lambda}\left\{\braket{(-\dt\phi - \sum_{i=1}^m \lambda_i G_i\delta_{\tau_i},- \Dx\phi,- \demi\Dxx\phi)}{(\rho,\cA,\cB)} - \int_{\R^d} \phi(0,x)\,d\mu_0 + \sum_{i=1}^m \lambda_i c_i \right\}.
}
As proved in Lemma \ref{lemma:A2}, the objective $\cV$ can be expressed as
\eqn{
  \cV=\inf_{(\rho,\cA,\cB)\in\cM(\Lambda,\cX)} (\Phi^* + \Psi^*)(\rho,\cA,\cB) = \inf_{(\rho,\cA,\cB)\in C_b(\Lambda,\cX)^*}(\Phi^* + \Psi^*)(\rho,\cA,\cB).
}

Let $O^{m\times n}$ denote a null matrix of size $m\times n$. Consider the point $(r,a,b)=(-1,O^{d\times 1}, O^{d\times d})$ which can be represented by $(\phi,\lambda) = (T-t,O^{m\times 1})$. As $F$ is nonnegative, at $(-1,O^{d\times 1}, O^{d\times d})$ we have 
\eqn{
  -1+F^*(O^{d\times 1}, O^{d\times d})=-1-\inf_{\alpha\in \R^d,\beta\in\bS^d} F(\alpha,\beta)< 0.
}
This shows that
\eqn{
  \Phi(-1,O^{d\times 1}, O^{d\times d}) = 0,\quad \Psi(-1,O^{d\times 1}, O^{d\times d}) = 0.
}
Thus, at $(-1,O^{d\times 1}, O^{d\times d})$, $\Phi$ is continuous with respect to the uniform norm (since $F^*$ is continuous in $\operatorname{dom}(F^*)$), and $\Psi$ is finite. Furthermore, as the convex functionals $\Phi$ and $\Psi$ take values in $(-\infty,+\infty]$, all of the required conditions are fulfilled to apply the Fenchel--Rockafellar duality theorem \citep[see e.g.,][Chapter 1]{brezis2011}. We then obtain
\eqn{
  \cV = \inf_{(\rho,\cA,\cB)\in C_b(\Lambda,\cX)^*}\{ \Phi^*(\rho,\cA,\cB) + \Psi^*(\rho,\cA,\cB) \} = \sup_{(r,a,b)\in C_b(\Lambda,\cX)}\{ -\Phi(-r,-a,-b) - \Psi(r,a,b) \},
}
and the infimum is in fact attained. Consequently, 
\eqn{
  \cV = \sup_{(r,a,b)} \bigg\{ -\int_{\R^d} \phi(0,x)\,d\mu_0 + \sum_{i=1}^m \lambda_i c_i \;;\; -r+F^*(-a,-b)\leq 0 \bigg\},
}
where the supremum is restricted to all $(r,a,b)$ represented by $(\phi,\lambda)\in BV([0,T],C_b^2(\R^d))\times\R^m$. Writing $(r,a,b)$ in terms of $(\phi,\lambda)$ with $\phi(T,\cdot)=0$, we obtain the required result.
\end{proof}

\subsection{Viscosity solutions}\label{sec:viscosity}

Adopting the concept of viscosity solutions, it can be shown that the supremum of the objective with respect to $\phi$ is achieved by the viscosity solution of the HJB equation \eqref{eq:pde_dual}. Due to presence of the Dirac delta functions in \eqref{eq:pde_dual}, we shall introduce a suitable definition of the viscosity solution that allows to have jump discontinuities in time. 
\begin{definition}\label{def:set}
  Denote by $\operatorname{set}(\tau)$ the set of entries of vector $\tau$ and by $K$ the cardinality of $\operatorname{set}(\tau)$. Let $t_0=0$, we define disjoint intervals $I_k:=[t_{k-1},t_{k})$ such that
  \eqn{
     \bigcup_{k=1}^{K}I_k = [0, T),
  }
  where $t_{k-1}<t_{k}$ and $t_k\in\operatorname{set}(\tau)$ for all $k=1,\ldots,K$.
\end{definition}

\begin{definition}[Viscosity solution]
  For any $\lambda\in\R^m$, we say $\phi$ is a viscosity subsolution (resp., supersolution) of \eqref{eq:pde_dual} if $\phi$ is a classical (continuous) viscosity subsolution (resp., supersolution) of (\ref{eq:pde_dual}) in $I_k\times\R^d$ for all $k=1,\ldots,K$, and has jump discontinuities:
  \eqn{
    \phi(t,x) = \phi(t^-,x) - \sum_{i=1}^m \lambda_i G_i(x){\mathds 1}(t=\tau_i) \qquad \forall (t, x)\in\tau\times\R^d.
  }
  Then, $\phi$ is called a viscosity solution of (\ref{eq:pde_dual}) if $\phi$ is both a viscosity subsolution and a viscosity supersolution of (\ref{eq:pde_dual}).
\end{definition}

\begin{remark}[Comparison principle]\label{remark:comparison}
  The comparison principle still holds for viscosity solutions of (\ref{eq:pde_dual}). Let $u$ and $v$ be a viscosity subsolution and a viscosity supersolution of the equation (\ref{eq:pde_dual}), respectively. At the terminal time $T$, $u(T,\cdot)\leq v(T,\cdot)$. Since $t_K=T$ is in $\operatorname{set}(\tau)$ and $u, v$ have the same jump size at $\{T\}\times\R^d$, we get $u(T^-, \cdot)\leq v(T^-, \cdot)$. Next, in the interval $I_K=[t_{K-1},t_K)$, by the classical comparison principle, we get $u\leq v$ on $I_K$. Applying this argument for all intervals $I_k$ for $k=1,\ldots,K$, we conclude that
  \eqn{
    u(t,x)\leq v(t,x), \quad \forall (t,x)\in[0,T]\times \R^d.
  }
  Also, $u(0,\cdot) \leq v(0,\cdot)$.
\end{remark}

\begin{remark}[Existence and uniqueness]\label{remark:existence}
  As a consequence of the comparison principle, there exists a unique viscosity solution of (\ref{eq:pde_dual}). The uniqueness is a direct consequence of the comparison principle. The existence can be obtained by Perron's method \citep[see][]{Pierre1992userguide} under which the comparison principle is a key argument.
\end{remark}

Now we shall prove Proposition \ref{prop:optimal_sol2}. The proof relies on a smoothing argument used in \citet{bouchard2017impact}, which is based on the shaken coefficients technique of \citet{Krylov2000shaken}. The proof is similar to Theorem 2.4 in \citet{bouchard2017impact}, which we sketch here for completeness.
\begin{proof}[Proof of Proposition \ref{prop:optimal_sol2}]
  Denote by $\varphi$ a viscosity solution of the equation (\ref{eq:pde_dual}) with any $\lambda\in\R^m$. From Remark \ref{remark:existence}, we know that such $\varphi$ exists and is unique. The first part of the proposition is proved in two steps:
  ~\\\bbf{Step 1.} Assuming that there exists a sequence of supersolutions of (\ref{eq:pde_dual}) in $BV([0,T],C_b^2(\R^d))$ converging to $\varphi$ pointwise, we can show that $\varphi$ achieves the supremum with respect to $\phi$ in the objective of the dual (\ref{eq:dual_obj_thm}). Let $\phi\in BV([0,T],C_b^2(\R^d))$ be any solution that satisfies (\ref{ieq:hjb}), and $\phi$ is also a (viscosity) supersolution of (\ref{eq:pde_dual}). By Remark \ref{remark:comparison}, we have $\varphi(0,x)\leq \phi(0,x)$ for all $x\in\R^d$, hence
  \eq{\label{ieq:supersolution}
    \sum_{i=1}^m \lambda_i c_i - \int_{\R^d} \phi(0,x)\,d\mu_0 \leq \sum_{i=1}^m \lambda_i c_i - \int_{\R^d} \varphi(0,x)\,d\mu_0.
  }
  The equality can be achieved in (\ref{ieq:supersolution}) by taking the supremum with respect to $\phi$ on the left-hand side of (\ref{ieq:supersolution}).
  ~\\\bbf{Step 2.} Now, we shall construct the sequence of supersolutions required in Step 1. Let us introduce the regularising kernel $r_\varepsilon:\R^d\to\R$ such that $r_\varepsilon(x) = \frac{1}{\varepsilon^d}r'\left(\frac{x}{\varepsilon}\right)$ where $r'$ is some compactly supported function that satisfies $\int_{\R^d} r'(x)\,dx=1$. Then we define $\varphi_\varepsilon = \varphi * r_\varepsilon$ where the convolution acts only on the variable $x$. By applying the result of \citet{bouchard2017impact} which relies critically on the fact that $F^*(a,b)$ is convex in $(a,b)$, it can be shown that $\varphi_\varepsilon$ are supersolutions of equation (\ref{eq:pde_dual}). If we send $\varepsilon$ to $0$, the supersolutions $\varphi_\varepsilon$ converge to the viscosity solution $\varphi$ pointwise. The desired sequence is then constructed.
  
  Now we prove the second part of the proposition. Let $(\rho^*,\alpha^*,\beta^*)$ be the optimal solution of Problem \ref{prob:pde}, then $(\rho^*,\rho^*\alpha^*,\rho^*\beta^*)$ also achieves the infimum \eqref{eq:cV_infsup}. Assume that there exists an optimal solution $\lambda^*\in\R^m$ that solves \eqref{eq:dual_obj}, then $(\phi_{\lambda^*},\lambda^*)$ also achieve the supremum in \eqref{eq:cV_infsup}. With the optimal solutions defined above, we can reformulate \eqref{eq:cV_infsup} as
  \eqn{
    0 &= \int_0^T\int_{\R^d} \left( F(\alpha^*,\beta^*) -\dt\phi_{\lambda^*} - \Dx\phi_{\lambda^*}\cdot \alpha^* - \demi\Dxx\phi_{\lambda^*}:\beta^* - \sum_{i=1}^m \lambda^*_i G_i\delta_{\tau_i} \right)d\rho^*_t dt \\
      &= \int_0^T\int_{\R^d} \left( F(\alpha^*,\beta^*) + F^*\left(\Dx\phi_{\lambda^*},\demi\Dxx\phi_{\lambda^*} \right) - \Dx\phi_{\lambda^*}\cdot \alpha^* - \demi\Dxx\phi_{\lambda^*}:\beta^* \right)d\rho^*_t dt .
  }
  Let $(\tilde\alpha,\tilde\beta)$ be defined by
  \eqn{
    (\tilde\alpha,\tilde\beta) = \nabla F^*\left(\Dx\phi_{\lambda^*},\demi\Dxx\phi_{\lambda^*} \right),\quad \left(\Dx\phi_{\lambda^*},\demi\Dxx\phi_{\lambda^*}\right)=\nabla F(\tilde\alpha,\tilde\beta).
  }
  Hence, by the definition of convex conjugate and the strong convexity of $F$,
  \eqn{
    0 &= \int_0^T\int_{\R^d} \left( F(\alpha^*,\beta^*) - F(\tilde\alpha,\tilde\beta) - \Dx\phi_{\lambda^*}\cdot (\alpha^*-\tilde\alpha) - \demi\Dxx\phi_{\lambda^*}:(\beta^*-\tilde\beta) \right)d\rho^*_t dt \\
      &\geq \int_0^T\int_{\R^d} C\left( \norm{\alpha^*-\tilde\alpha}^2 + \norm{\beta^*-\tilde\beta}^2 \right) d\rho^*_t dt \geq 0,
  }
  where $C>0$ is a constant. Therefore, $(\alpha^*,\beta^*)=(\tilde\alpha,\tilde\beta)$, $d\rho^*_t dt$-almost everywhere. The proof is completed.
\end{proof}

\section{LSV Calibration}

In this section, we illustrate our method by calibrating a Heston-like LSV model. This method could also be easily extended to other LSV models. We consider the LSV model with following dynamics under the risk-neutral measure:
\eq{\label{eq:LSV}
  \left\{ \begin{array}{l}
  dZ_t = (r(t) - q(t) - \demi\sigma^2(t,Z_t,V_t))\,dt + \sigma(t,Z_t,V_t)\,dW_t^Z, \\
  dV_t = \kappa(\theta - V_t)\,dt + \xi\sqrt{V_t}\,dW_t^V, \\
  dW_t^Z dW_t^V = \eta(t,Z_t,V_t) \,dt,
  \end{array} \right.
}
where $Z_t$ is the logarithm of the stock price at time $t$. The interpretations of $r$ and $q$ differ between financial markets. In the equity market, $r$ is the risk-free rate and $q$ is the dividend yield. In the FX market, $r$ is the domestic interest rate and $q$ is the foreign interest rate. The parameters $\kappa, \theta, \xi$ have the same interpretation as in the Heston model. In our method, we assume these parameters are given and obtained by calibrating a pure Heston model. Note that in the literature, the widely considered LSV model has a volatility function $\sigma(t,Z_t,V_t)=L(t,Z_t)\sqrt{V_t}$ and a constant correlation $\eta$, where $L(t,Z_t)$ is known as the leverage function. By contrast, we consider a local-stochastic volatility $\sigma>0$ and a local-stochastic correlation $\eta\in[-1,1]$ whose values depend on $(t,Z_t,V_t)$. Our objective is to calibrate $\sigma(t,Z,V)$ and $\eta(t,Z,V)$ so that model prices exactly match market prices.

\begin{remark}\label{rmk:leverage}
If the volatility $\sigma(t,Z,V)\equiv \sqrt{V}$ and the correlation $\eta(t,Z,V)$ is a constant, the LSV model reduces to a pure Heston model. Furthermore, if $\sigma(t,Z,V)$ is independent of the variable $V$, the model is equivalent to a local volatility model.
\end{remark}

Consider a probability measure $\bP\in\cP^1$ and a two-dimensional $\bP$-semi-martingale $X_t$. The process $X_t$ has dynamics (\ref{eq:LSV}), i.e., $X_t=(Z_t,V_t)$, if $\bP$ is characterised by $(\alpha^\bP,\beta^\bP)$ such that
\eq{\label{eq:alpha_beta}
  (\alpha^\bP_t,\beta^\bP_t) = \left(\left[ \begin{array}{c}
r_t-q_t-\demi\sigma^2_t \\ \kappa(\theta - V_t)\end{array} \right],\,\left[ \begin{array}{cc}
\sigma^2_t & \eta_t\xi\sqrt{V_t}\sigma_t \\ \eta_t\xi\sqrt{V_t}\sigma_t & \xi^2V_t \end{array} \right] \right), \quad t\in[0,T],
}
with functions $\sigma_t=\sigma(t,Z_t,V_t)$ and $\eta_t=\eta(t,Z_t,V_t)$. Recall that the parameters $(\kappa, \theta,\xi)$ are assumed to be given. Also, $r_t$ and $q_t$ are known and $V_t$ is a state variable. Hence, the only unknown variables in \eqref{eq:alpha_beta} are $\sigma_t$ and $\eta_t$. As we will see below, $\sigma_t$ will be the only free variable in the calibration. Given $m$ European options with prices $c\in\R_+^m$, maturities $\tau=(\tau_1,\ldots,\tau_m)\in(0,T]^m$ and discounted payoffs $G = (G_1,\ldots,G_m)$ where $G_i:\R^2\to\R_+$ (e.g., $G_i(x)=e^{-\int_0^{\tau_i}r(s)ds}(e^{x_1}-K)^+$ if the $i$-th option is a European call with strike $K$ and maturity $\tau_i$, where $x_1$ stands for the first element of $x$). If $X_t$ has an initial distribution $\mu_0=\delta_{(Z_0,V_0)}$ and is exactly calibrated to these European options, then $\bP\in\cP(\mu_0,\tau,c,G)$. One way to build a calibrated LSV model is to solve
\eq{\label{eq:main_cali}
  \cV = \inf_{\bP\in\cP(\mu_0,\tau,c,G)}\EP \int_0^T F(t,X_t,\alpha_t^\bP,\beta_t^\bP) \,dt,
}
where $F$ is a suitable convex cost function that forces $(\alpha^\bP,\beta^\bP)$ to take the form of (\ref{eq:alpha_beta}).

One possible way to choose the cost function $F$ is based on the idea of minimising the difference between each element of $\beta^\bP$ and a reference value while keeping $\beta^\bP$ in $\bS_+^2$. However, it is often impossible to find an explicit formula to approximate $F^*$. Thus numerical optimisation is needed, which makes the method computationally expensive. To overcome this issue, we choose the correlation
\eq{\label{eq:eta}
\eta_t = \frac{\sqrt{V_t}}{\sigma_t}\etabar, \quad t\in[0,T],
}
where $\etabar$ is a constant correlation obtained (along with $\kappa, \theta, \xi$) by calibrating a pure Heston model. In this case, $\beta^\bP_t$ is positive semidefinite if and only if $\sigma^2_t\geq \etabar^2 V_t$ for $t\leq T$.

\begin{figure}[H]
  \begin{center}
  \includegraphics[width=0.8\textwidth]{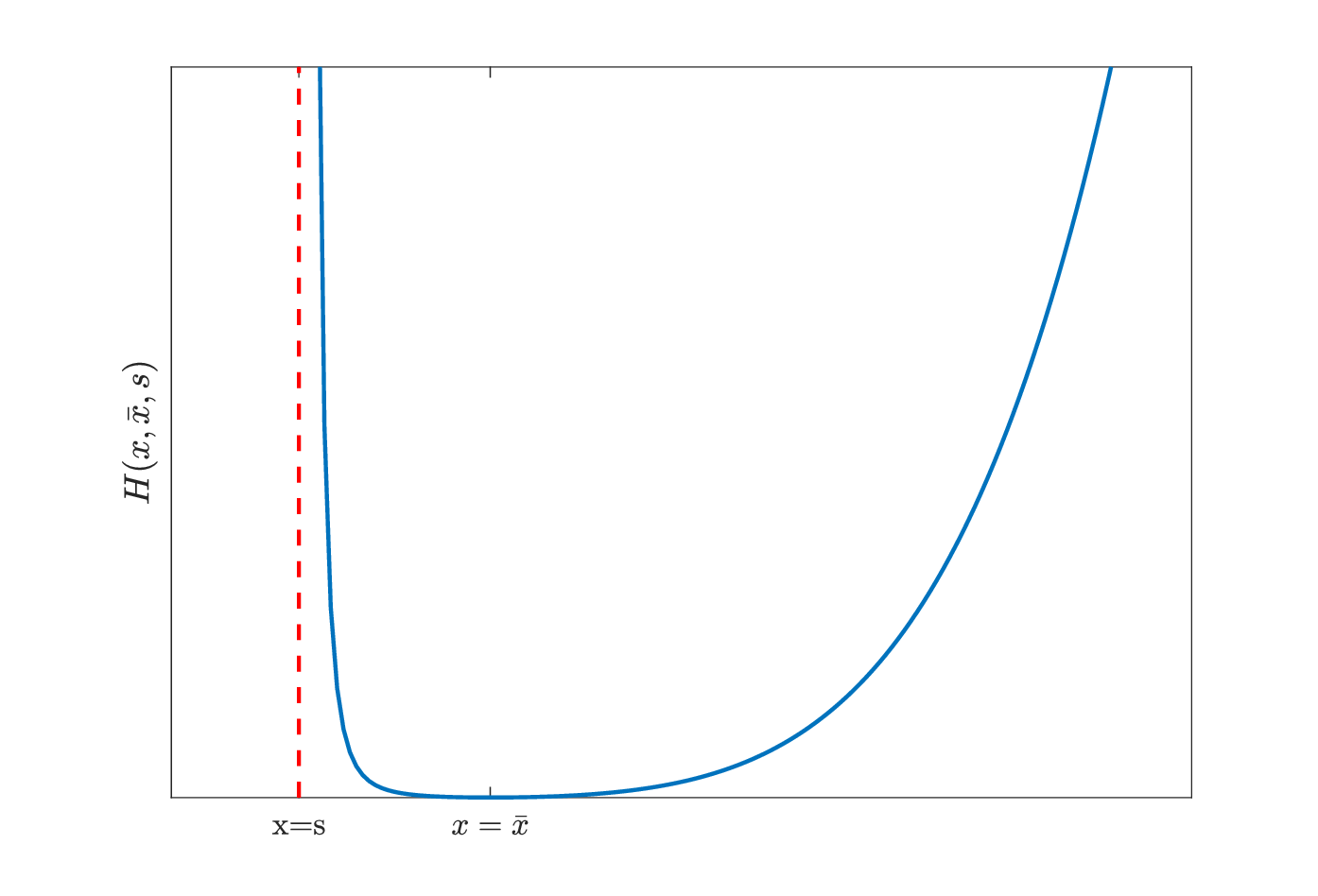}
  \caption{The function $H(x,\xbar,s)$ for a given $\xbar$ and a given $s<\xbar$.}
  \label{fig:funch}
  \end{center}
\end{figure}

\begin{definition}
  Define function $H:\R\times\R_+\times\R\to\R\cup\{+\infty\}$ such that
  \eqn{
    H(x,\xbar,s):= \left\{ \begin{array}{ll}\displaystyle a(\frac{x-s}{\xbar-s})^{1+p} + b(\frac{x-s}{\xbar-s})^{1-p} + c  & \mbox{if } x > s \mbox{ and } \xbar > s, \\
			   +\infty & \mbox{otherwise.} \end{array} \right.
  }
  The parameter $p$ is a constant greater than 1, and $a,b,c$ are constants determined to minimise the function at $x=\xbar$ with $\min H=0$.
\end{definition}
Given $\xbar$ and $s$ satisfying $\xbar>s$, the function $H$ is convex in $x$ and minimised at $\xbar$. It is finite only when $x>s$. In the numerical examples (see Section \ref{sec:num_exp_sim} and \ref{sec:num_exp_market} below), the parameter $p$ is set to $4$. A plot of $H$ is given in Figure \ref{fig:funch}. Then, we define the cost function as follows.
\begin{definition}
  The cost function $F:\R\times\R\times\R\times\R^2\times\bS^2\to\R\cup\{+\infty\}$ is defined as
  \eq{\label{eq:LSV_costfun}
    F(t,Z,V,\alpha,\beta) := \left\{ \begin{array}{ll} H(\beta_{11},V, \etabar^2 V)  & \mbox{if } (\alpha,\beta)\in\Gamma(t,V), \\
			 +\infty & \mbox{otherwise,} \end{array} \right.
  }
  where the convex set $\Gamma$ is defined as
  \eqn{
    \Gamma(t,V) := \{ (\alpha,\beta)\in\R^2\times\bS^2 \mid  \alpha_1=r(t)-q(t)-\beta_{11}/2,\, \alpha_2=\kappa(\theta-V),\, \beta_{12}=\beta_{21}=\etabar\xi V,\, \beta_{22}=\xi^2V \} .
  }
\end{definition}

\begin{remark}
  The function $H$ penalises deviations of the LSV model from a pure Heston model by choosing $\xbar=V$ (see Remark \ref{rmk:leverage}). This approach seeks to retain the attractive features of the Heston model while still matching all the market prices. We also set $s=\etabar^2 V$ to ensure that $\sigma^2>\etabar^2 V$, hence $\beta$ remains positive definite and the correlation $\eta$ is in $[-1,1]$. 
  The set $\Gamma$ forces $X_t$ to have dynamics of the form \eqref{eq:LSV} with $\eta$ defined in \eqref{eq:eta} by restricting the characteristics in $\Gamma$. In particular, it remains risk neutral.
\end{remark}

By applying Proposition \ref{prop:optimal_sol2}, the dual formulation of (\ref{eq:main_cali}) with the cost function \eqref{eq:LSV_costfun} is as follows:
\eq{\label{eq:lsv_dual_obj}
  \cV = \sup_{\lambda\in\R^m}\left\{ \sum_{i=1}^m\lambda_i c_i - \phi_\lambda(0,Z_0,V_0) \right\},
}
where $\phi_\lambda$ is the viscosity solution to the HJB equation
\eq{\label{eq:lsv_hjb}
\begin{split}
\dt\phi_\lambda + \sum_{i=1}^m\lambda_i G_i\delta_{\tau_i} &+ \sup_{\beta_{11}}\bigg\{ (r-q-\demi\beta_{11})\dz\phi_\lambda + \kappa(\theta-V)\dv\phi_\lambda  + \etabar\xi V\dzv\phi_\lambda \\
& + \demi\beta_{11}\dzz\phi_\lambda + \demi\xi^2V\dvv\phi_\lambda - H(\beta_{11},V, \etabar^2 V) \bigg\} = 0,
\end{split}
}
with a terminal condition $\phi_\lambda(T,\cdot) = 0$. 

Given any $\lambda\in\R^m$, we can calculate $\phi_\lambda(0,Z_0,V_0)$ by numerically solving the HJB equation (\ref{eq:lsv_hjb}). The optimal $\lambda$ can be found through a standard optimisation algorithm (see Section \ref{sec:num_method} below). The convergence of the algorithm can be improved by providing the gradient of the objective. Let $\bbeta_\lambda$ denote the optimal $\beta_{11}$ that solves the supremum in \eqref{eq:lsv_hjb}, which also implicitly depends on $\lambda$. In fact, solving the supremum in \eqref{eq:lsv_hjb} is equivalent to solving the following equation for $\sigma^2$:
\eq{\label{eq:lsv_optimal_sigma}
  (\dzz\phi_\lambda - \dz\phi_\lambda)/2 = \partial_{\sigma^2}H(\sigma^2,V, \etabar^2 V),
}
for which a closed-form solution is available. We also denote by $\bP_\lambda\in\cP^1$ a probability measure characterised by $(\alpha^{\bP_\lambda}, \beta^{\bP_\lambda})$ defined in \eqref{eq:alpha_beta} with $\displaystyle (\sigma_t,\eta_t)=(\sqrt{(\bbeta_\lambda)_t}, \etabar\sqrt{V_t/(\bbeta_\lambda)_t}),\, t\leq T$.
\begin{lemma}
  Define $J(\lambda) = \sum_{i=1}^m\lambda_i c_i - \phi_\lambda(0,Z_0,V_0)$. The gradient of $J(\lambda)$ with respect to $\lambda_i$ can be formulated as:
  \eq{\label{eq:lsv_gradient}
    \partial_{\lambda_i}J(\lambda) = c_i - \E^{\bP_\lambda} G_i(X_{\tau_i}), \quad \forall i=1,\ldots,m.
  }
  In addition, $\E^{\bP_\lambda}G_i(X_{\tau_i})=\phi'(0,Z_0,V_0)$ where $\phi'$ solves
  \eq{\label{eq:pricing_PDE_alt}
    \dt\phi' + (r-q-\demi\bbeta_\lambda)\dz\phi' + \kappa(\theta-V)\dv\phi'  + \etabar\xi V\dzv\phi' + \demi\bbeta_\lambda\dzz\phi' + \demi\xi^2V\dvv\phi' = 0,
  }
  with the terminal condition $\phi'(\tau_i,\cdot)=G_i$.
\end{lemma}
\begin{proof}
  Given a $\lambda$ and the associated $\bbeta_\lambda$, the HJB equation \eqref{eq:lsv_hjb} reduces to
  \eq{\label{eq:reduced_lsv_hjb}
  \begin{split}
  \dt\phi_\lambda + \sum_{i=1}^m\lambda_i G_i\delta_{\tau_i} &+ (r-q-\demi\bbeta_\lambda)\dz\phi_\lambda + \kappa(\theta-V)\dv\phi_\lambda  + \etabar\xi V\dzv\phi_\lambda \\
  & + \demi\bbeta_\lambda\dzz\phi_\lambda + \demi\xi^2V\dvv\phi_\lambda - H(\bbeta_\lambda,V, \etabar^2 V)  = 0.
  \end{split}
  }
  Since $\lambda,\phi_\lambda$ and $\bbeta_\lambda$ are related implicitly, by taking implicit partial differentiation of \eqref{eq:reduced_lsv_hjb} to compute $\phi':=\partial_{\lambda_i}\phi_\lambda$ for any $i=1,\ldots,m$, we obtain the following PDE
  \eq{\label{eq:pricing_PDE}
  \begin{split}
    \dt\phi' + (r-q-\demi\bbeta_\lambda)\dz\phi' &+ \kappa(\theta-V)\dv\phi'  + \etabar\xi V\dzv\phi' \\
	&+ \demi\bbeta_\lambda\dzz\phi' + \demi\xi^2V\dvv\phi' = -G_i\delta_{\tau_i}.
  \end{split}
  }
  With the terminal condition $\phi'(T,\cdot) = 0$, (\ref{eq:pricing_PDE}) can be solved by the Feynman--Kac formula \citep[see e.g.,][Theorem 7.6]{shreve1991calculus}. Thus,
  \eqn{
    \phi'(0,Z_0,V_0)=\E^{\bP_\lambda} G_i(X_{\tau_i}).
  }
  Moreover, solving \eqref{eq:pricing_PDE} with $\phi'(T,\cdot) = 0$ is equivalent to solving \eqref{eq:pricing_PDE_alt} with $\phi'(\tau_i,\cdot)=G_i$. The proof is completed.
\end{proof}

\begin{remark}
  Note that $\E^{\bP_\lambda} G_i(X_{\tau_i})$ is the price of the $i$-th European option calculated by $X_t$ under $\bP_\lambda$, which we refer to as the model price, and $c_i$ is the market price. Instead of solving \eqref{eq:pricing_PDE} once for each option, we can perform a Monte Carlo simulation to efficiently calculate the model prices for all options. However, for the sake of accuracy, we still choose to solve \eqref{eq:pricing_PDE} in the numerical examples below. Moreover, as the gradient is decreasing to zero while the solution is moving towards the optimal solution, the optimisation process can be interpreted as matching the model $X_t$ to market prices.
\end{remark}

\section{Numerical aspects}
\subsection{Numerical method}\label{sec:num_method}

In this section, we present a numerical method for solving the dual formulation. To shorten notations, we will simply write $\phi$ for $\phi_\lambda$ from now on. Starting with an initial $\lambda=\lambda^0$ (e.g., setting it to a null vector), we solve the HJB equation \eqref{eq:lsv_hjb} to calculate $\phi(0,Z_0,V_0)$ and hence calculate $J(\lambda^0)$. Then, $J$ is maximised over $\lambda\in\R^m$ through an optimisation algorithm. In particular, we employed the L-BFGS algorithm \citep{LBFGS1989} and obtained good convergence. The optimisation process can be accelerated by providing the gradient $\nabla J(\lambda)$ which can be numerically computed by (\ref{eq:lsv_gradient}). We measure the optimality by the maximum absolute value on the gradient. In other words, by setting a threshold $\epsilon_1$, the algorithm terminates when the following stopping criterion is reached:
\eqn{
  \norm{\nabla J(\lambda)}_\infty \leq \epsilon_1.
}

For solving the HJB equation \eqref{eq:lsv_hjb}, we use an alternating direction implicit (ADI) method together with the central finite difference scheme. In the numerical examples below, we employ the Douglas scheme from \citet{foulon2010adi}. Given a $\lambda$, we solve the HJB equation backward. Consider a discretisation $\{t_k\}$ of the time interval $[0,T]$ such that $0=t_0<t_1<\cdots<t_{N_T}=T, N_T\in\N$. Without loss of generality, we assume that $\operatorname{set}(\tau)\subset \{t_k\}$. At each time step $t_k$, we approximate $\sigma_{t_k}^2$ by solving (\ref{eq:lsv_optimal_sigma}) with $\phi=\phi_{t_{k+1}}$ for which an analytical solution can be found. At $t=t_k$, with the approximated $\sigma_{t_k}^2$, the HJB equation \eqref{eq:lsv_hjb} is solved by the ADI finite difference method. Note that this approximation scheme of $\sigma$ is similar to the one used in \citet{ren2007calibrating} for approximating the leverage function. 

Let $\hat\tau_i$ be an element in $\operatorname{set}(\tau)$ such that $\cup_{i=1}^K\{\hat\tau_i\}=\operatorname{set}(\tau)$ and $0=:\hat\tau_0<\hat\tau_1<\ldots<\hat\tau_K=T$ (see Definition \ref{def:set} for the definitions of $\operatorname{set}(\tau)$ and $K$). Denote by $D$ the spatial computational domain and by $\partial D$ the boundary of $D$. When numerically solving the HJB equation \eqref{eq:lsv_hjb}, we impose the following boundary conditions for the spatial dimensions:
\eqn{
  \forall i&=1,\ldots,K  & \Dxx\phi(t,x)&=\Dxx\phi(\hat\tau_i^-,x), \qquad (t,x)\in[\hat\tau_{i-1}, \hat\tau_{i})\times\partial D,
}
In addition, we set a sufficiently large $D$ to reduce the impact of the boundary conditions. 

To handle the jump discontinuities caused by the presence of the Dirac delta terms, we can solve the HJB equation interval-wise in the intervals separated by the maturities, and the jump discontinuity can be incorporated into the terminal condition of the HJB equation in each interval. More precisely, if $t_{k+1}$ is equal to the maturity of any calibrating options, we incorporate the jump discontinuity by adding $\sum_{i=1}^m\lambda_i G_i{\mathds 1}(t_{k+1}=\tau_i)$ to $\phi_{t_{k+1}}$. The numerical method is summarised in Algorithm \ref{alg1}.

\begin{algorithm}[t!]
  \DontPrintSemicolon
  \SetNoFillComment
  \KwData{Market prices of European option}
  \KwResult{A calibrated OT-LSV model that matches all market prices}
  
  Set an initial $\lambda$\;
  \Do{$\norm{\nabla J(\lambda)}_\infty > \epsilon_1$}{
    \tcc{Solving the HJB equation}
    \For{$k = N_T-1,\ldots,0$}{
      \If{$t_{k+1}$ is equal to the maturity of any calibrating options.}{
        $\phi_{t_{k+1}} \gets \phi_{t_{k+1}} + \sum_{i=1}^m\lambda_i G_i{\mathds 1}(t_{k+1}=\tau_i)$\;
      }
      Approximate $\sigma_{t_k}^2$ by solving (\ref{eq:lsv_optimal_sigma}) with $\phi=\phi_{t_{k+1}}$\;
      Solve the HJB equation (\ref{eq:lsv_hjb}) by the ADI method at $t=t_k$\;
    }
    \tcc{Calculating model prices and gradient}
    Solve (\ref{eq:pricing_PDE_alt}) to calculate the model prices by the ADI method\;
    Calculate the gradient $\nabla J(\lambda)$ by (\ref{eq:lsv_gradient})\;
    Update $\lambda$ by the L-BFGS algorithm\;
  }
  \caption{LSV calibration}
  \label{alg1}
\end{algorithm}

Due to the non-linearity of the HJB equation, when the time step sizes are too large, it might not be accurate to simply approximate $\sigma_{t_k}^2$ by solving (\ref{eq:lsv_optimal_sigma}) with $\phi=\phi_{t_{k+1}}$ once per time step. Therefore, we slightly modify the algorithm by including an iterative step to improve the accuracy of the approximation of $\sigma_{t_k}^2$. In the literature, this iterative step is known as \iit{policy iteration}, see e.g., \citet{Ma2017monotone}. Specifically, at each time step $t_k$, we first approximate $\sigma_{t_k}^2$ by solving (\ref{eq:lsv_optimal_sigma}) with $\phi=\phi_{t_{k+1}}$. Next, we obtain $\phi_{t_k}$ by solving the HJB equation \eqref{eq:lsv_hjb} with $\sigma_{t_k}^2$, and then approximate $\sigma_{t_k}^2$ again by solving (\ref{eq:lsv_optimal_sigma}) with $\phi=\phi_{t_k}$. This process is repeated until $\phi_{t_k}$ converges. For completeness, the modified numerical method with policy iteration is summarised in Algorithm \ref{alg2}. For the sake of accuracy, we use Algorithm \ref{alg2} in both Section \ref{sec:num_exp_sim} and Section \ref{sec:num_exp_market}.

In our experiments, we notice that the algorithm can provide satisfactory calibration results even with coarse grids. However, it is crucial to ensure that the grids are fine enough, because we do not want to calibrate the wrong model prices to the calibrating option prices. In fact, we observe that the algorithm converges faster with finder grids, because the numerical approximations of the gradients are more accurate with finer grids.

\subsection{Numerical example: simulated data} \label{sec:num_exp_sim}

\begin{figure}[t!]
  \begin{center}
  \includegraphics[width=0.8\textwidth]{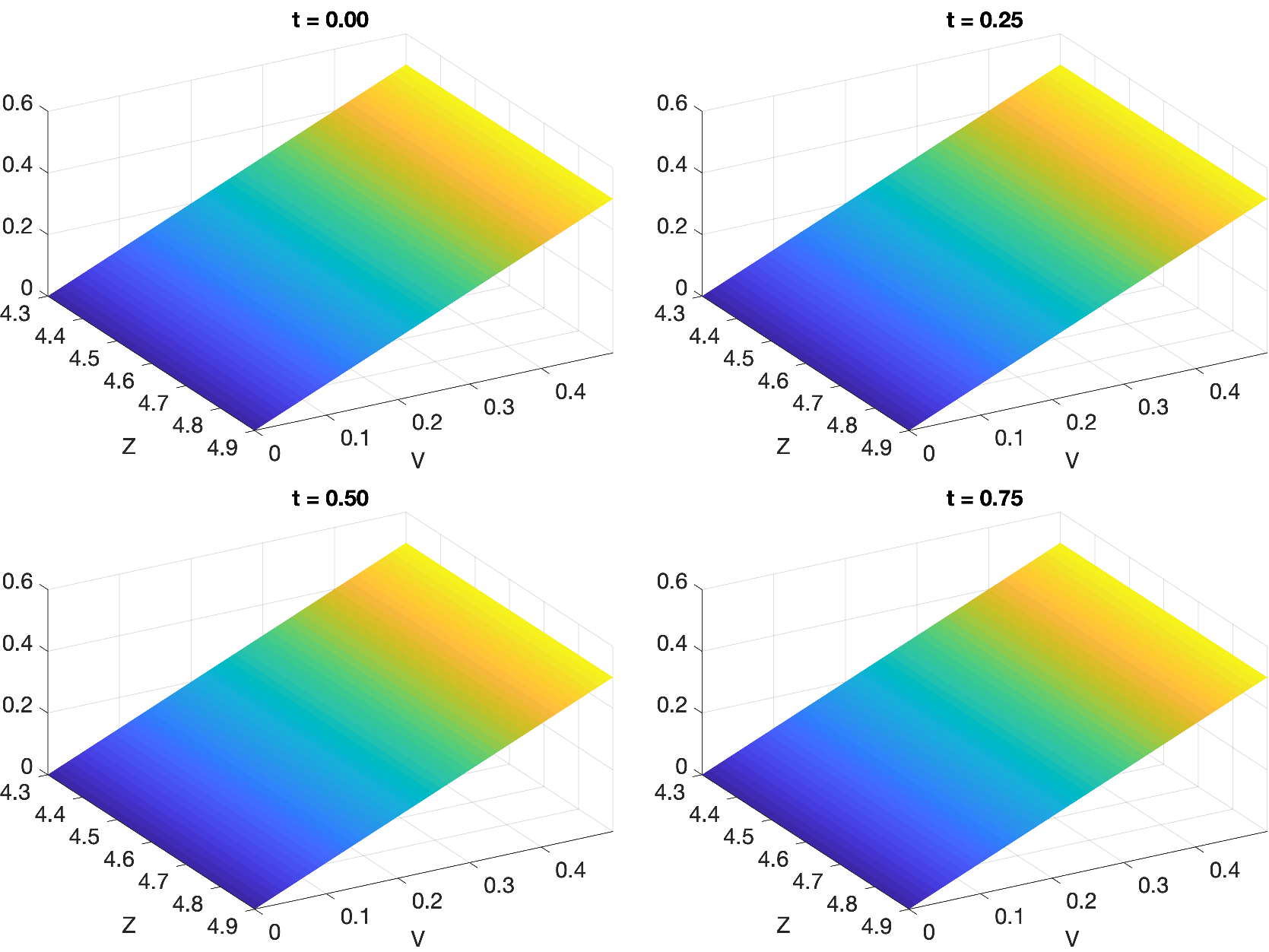}
  \caption{The volatility function $\sigma^2(t,Z,V)$ in Example 1}
  \label{fig:case1}
  \end{center}
\end{figure}

In this section, we provide two numerical examples with simulated data to demonstrate the calibration method. In both examples, the risk-free rate is set to a constant $r=0.05$ and the dividend yield is set to $q=0$. Let $Z_0 = \ln 100$ and $V_0 = 0.04$ for both models. We consider a uniform mesh over the spatial computational domain $D=[Z_0 - 4\sqrt{V_0}, Z_0 + 4\sqrt{V_0}]\times[0, 0.5]$ and use 101 points for each dimension. We also consider a uniform mesh over the time interval $[0,1]$ with $N_T=100$. The LSV model is calibrated to a set of European call options generated by a Heston model with given parameters. For clarity, we will refer to the LSV model as the \iit{OT-LSV model} and refer to the Heston model as the \iit{Heston generating model}. The option prices are calculated at maturities in $\{0.2, 0.4, 0.6, 0.8, 1.0\}$ and at 18 different strikes in $[Z_0 - 1.4\sqrt{V_0}, Z_0 + 1.4\sqrt{V_0}]$.

\subsubsection{Example 1}

In the first example, we use parameters $(\kappa,\theta,\xi,\etabar)=(0.5, 0.04, 0.16, -0.4)$ for both the OT-LSV model and the Heston generating model. This example represents a trivial case, since if we use the same set of parameters for both models, the optimal solution of the dual formulation is a null vector $\lambda=\mathbf{0}\in\R^m$, and hence $\cV = 0$. In this case, under the optimal measure of Problem \ref{prob:main}, $\sigma^2(t,Z,V)=V$ and $\eta(t,Z,V)=\etabar$. Setting a threshold $\epsilon_1 = 10^{-6}$, we obtain the expected results. The plot of $\sigma^2(t,Z,V)$ is provided in Figure \ref{fig:case1}.

\subsubsection{Example 2}

\begin{figure}[t!]
  \begin{center}
  \includegraphics[width=0.8\textwidth]{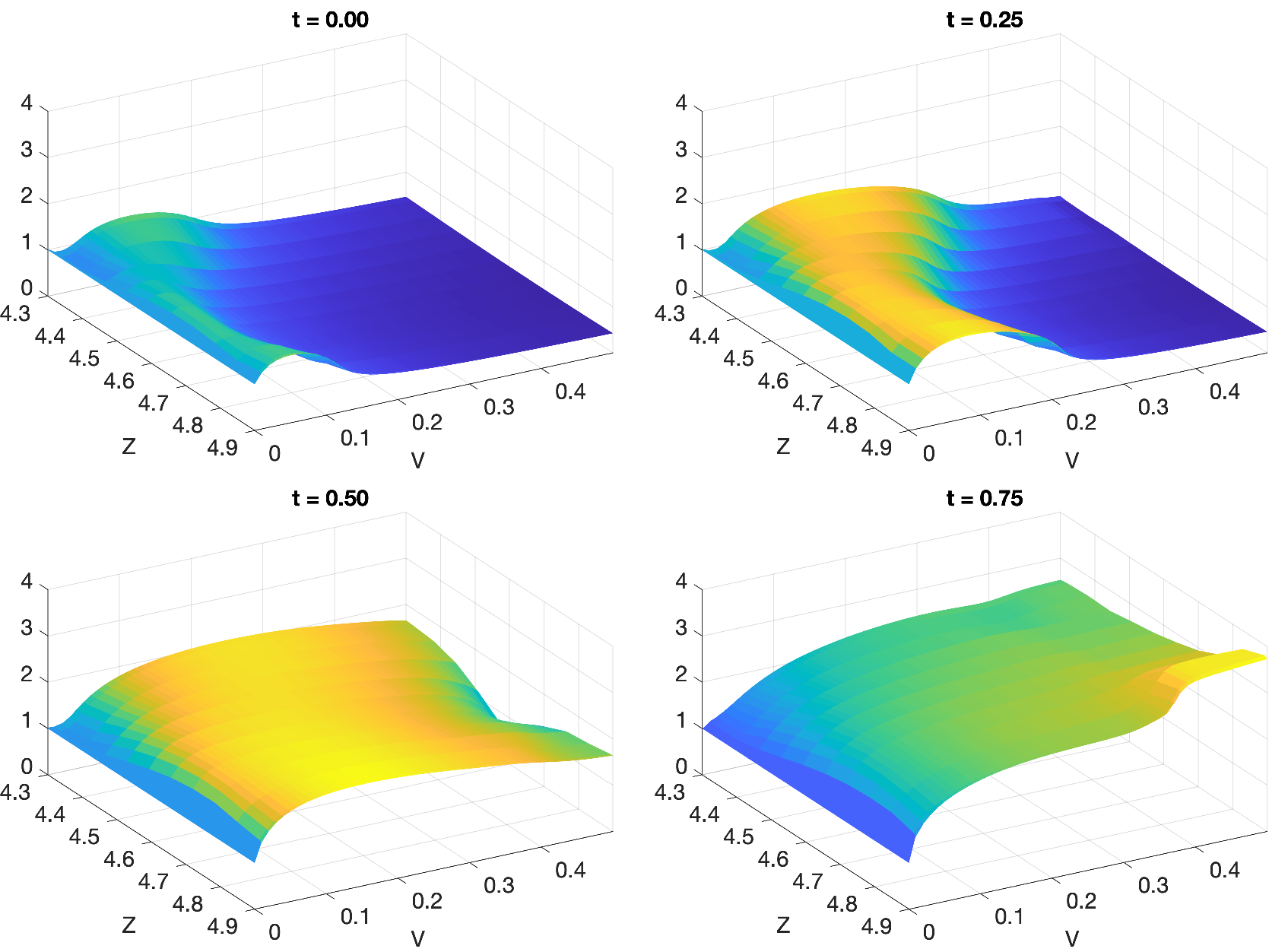}
  \caption{The function $\sigma^2(t,Z,V)/V$ in Example 2}
  \label{fig:case2}
  \end{center}
\end{figure}

In the second example, we give different parameters to the OT-LSV model and the Heston generating model (see Table \ref{table1}). As noted in Remark \ref{rmk:leverage}, the OT-LSV model reduces to a LV model if $\sigma^2(t,Z,V)$ is independent of $V$. Also, it is well known that an LV model can be calibrated to any arbitrage-free option prices. In this example, the Heston generating model has characteristics that are outside of $\Gamma$ in the cost function $F$, so the Heston generating model would lead to an infinite cost. However, since the generated option prices are arbitrage free, a finite cost is still achievable by the OT-LSV model and the problem is admissible, i.e., $\cP(\mu_0,\tau,c,G)\neq\emptyset$ and $\cV<+\infty$.

\vspace{1em}
\begin{table}[h]
  \begin{center}
  \begin{tabular}{|c|cccc|}
  \hline
         & $\kappa$ & $\theta$ & $\xi$ & $\etabar$ \\ \hline
  Heston generating model & 2.0 & 0.09 & 0.10 & -0.6 \\ \hline
  OT-LSV model  & 0.5 & 0.04 & 0.16 & -0.4 \\ \hline
  \end{tabular}
  \caption{The parameters of the Heston generating model and the OT-LSV model in Example 2}
  \label{table1}
  \end{center}
\end{table}
\vspace{-1em}

\begin{figure}[t!]
\begin{center}
\includegraphics[width=0.8\textwidth]{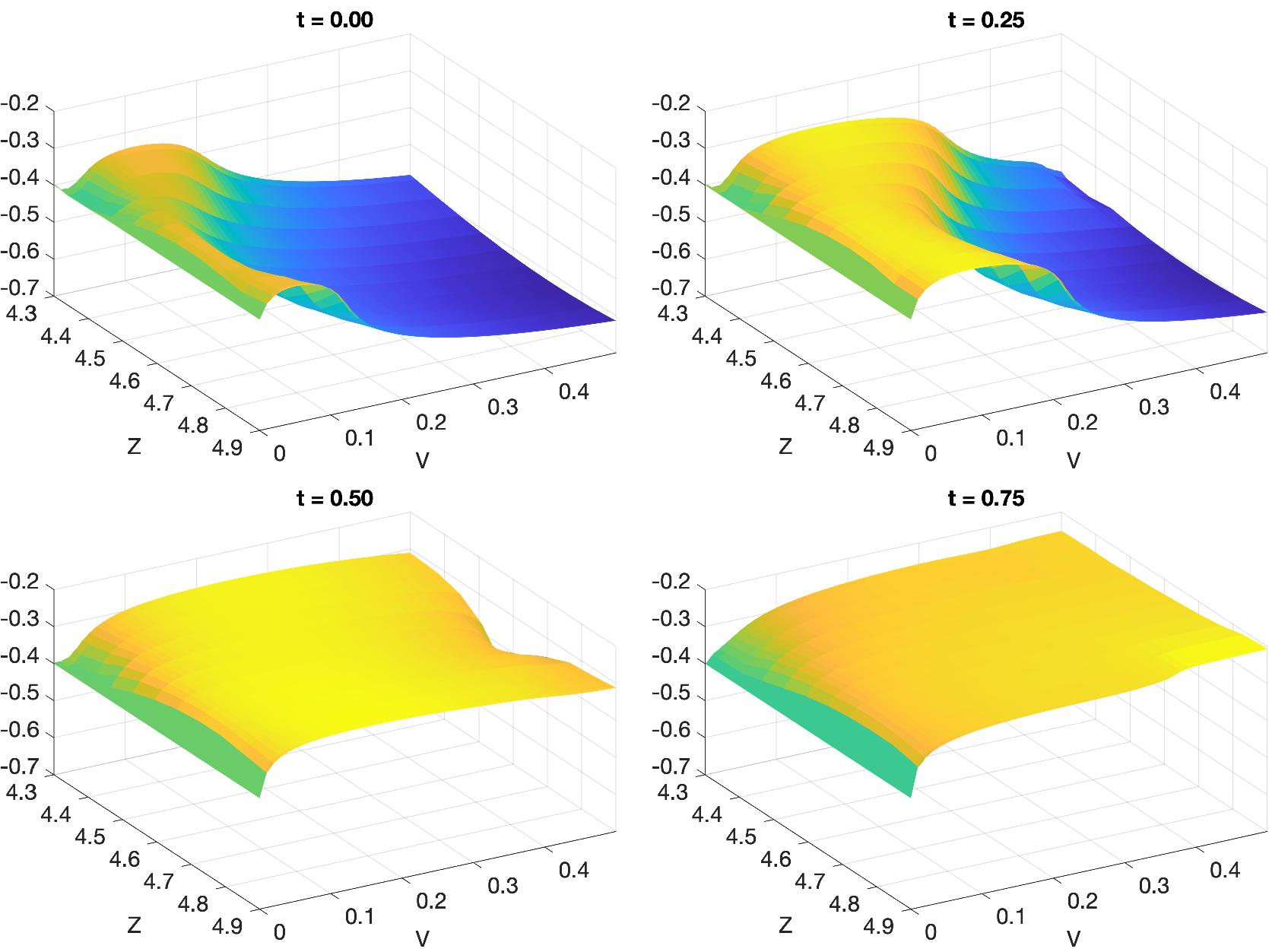}
\caption{The correlation function $\eta(t,Z,V)$ in Example 2}
\label{fig:case2eta}
\end{center}
\end{figure}

\begin{figure}[t!]
\begin{center}
\includegraphics[width=0.8\textwidth]{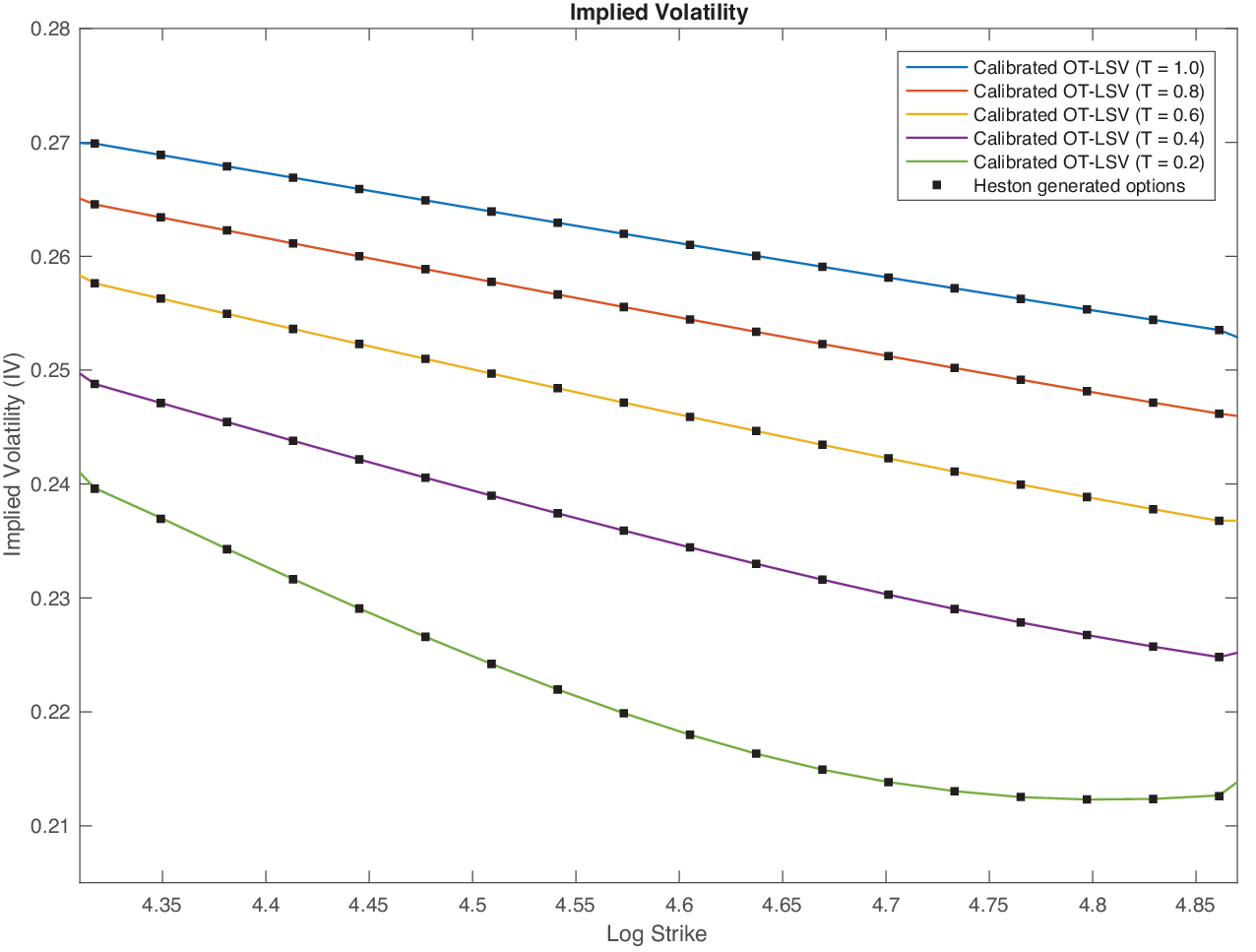}
\caption{The implied volatility of the Heston generated options and the calibrated OT-LSV model in Example 2}
\label{fig:case2_IV}
\end{center}
\end{figure}

By setting the threshold $\epsilon_1=0.0005$, we obtained accurate calibration results. The calibration results for a subset of options are given in Table \ref{table2}. If $\sigma^2$ is in the form of $\sigma^2(t,Z,V)=L^2(t,Z) V$ for some function $L$, then $L(t,Z)$ is called the leverage function and the OT-LSV model recovers the traditional LSV model considered in most of the literature. Thus, we plot the function $\sigma^2(t,Z,V)/V$ in Figure \ref{fig:case2} for comparison with $L^2(t,Z)$. The plot of the correlation function $\eta(t,Z,V)$ is also provided in Figure \ref{fig:case2eta}. Finally, we show the implied volatility of the Heston generated option prices and the OT-LSV generated option prices in Figure \ref{fig:case2_IV}. We can see that the OT-LSV model is well-calibrated to the Heston generated option prices. 

\begin{table}[b!]
  \small
  \begin{center}
  \begin{tabular}{|c|c|c|c|c|}
  \hline
  Maturity  & Log-strike  & Implied vol (Heston) & Implied vol (OT-LSV) & Error    \\ \hline
               & 4.3492     & 0.2396   & 0.2396   & 1.55E-05 \\
               & 4.4452     & 0.2291   & 0.2291   & 1.09E-06 \\
  T = 0.2      & 4.5732     & 0.2199   & 0.2199   & 8.89E-06 \\
               & 4.7012     & 0.2138   & 0.2138   & 8.56E-06 \\
               & 4.8292     & 0.2123   & 0.2124   & 2.99E-06 \\ \hline
               & 4.3492     & 0.2488   & 0.2488   & 1.82E-07 \\
               & 4.4452     & 0.2422   & 0.2422   & 3.93E-06 \\
  T = 0.4      & 4.5732     & 0.2359   & 0.2359   & 2.03E-06 \\
               & 4.7012     & 0.2303   & 0.2303   & 2.69E-06 \\
               & 4.8292     & 0.2257   & 0.2257   & 5.20E-07 \\ \hline
               & 4.3492     & 0.2576   & 0.2576   & 8.15E-06 \\
               & 4.4452     & 0.2523   & 0.2523   & 2.14E-07 \\
  T = 0.6      & 4.5732     & 0.2471   & 0.2471   & 2.42E-06 \\
               & 4.7012     & 0.2423   & 0.2423   & 6.52E-07 \\
               & 4.8292     & 0.2378   & 0.2378   & 3.55E-06 \\ \hline
               & 4.3492     & 0.2646   & 0.2646   & 1.97E-05 \\
               & 4.4452     & 0.2600   & 0.2600   & 1.82E-06 \\
  T = 0.8      & 4.5732     & 0.2555   & 0.2555   & 2.72E-06 \\
               & 4.7012     & 0.2512   & 0.2512   & 1.81E-06 \\
               & 4.8292     & 0.2472   & 0.2472   & 2.13E-06 \\ \hline
               & 4.3492     & 0.2699   & 0.2699   & 4.08E-06 \\
               & 4.4452     & 0.2659   & 0.2659   & 6.81E-07 \\
  T = 1.0      & 4.5732     & 0.2620   & 0.2620   & 1.44E-06 \\
               & 4.7012     & 0.2581   & 0.2581   & 1.54E-06 \\
               & 4.8292     & 0.2544   & 0.2544   & 7.30E-07 \\ \hline
  \end{tabular}
  \caption{A subset of the implied volatility of the options generated by the Heston generating model and the calibrated OT-LSV model in Example 2}
  \label{table2}
  \end{center}
\end{table}

\newpage
\subsection{Numerical example: FX market data} \label{sec:num_exp_market}

In this example, we calibrate the OT-LSV model to the FX options data provided in \citet{tian2015calibrating}. The options data and the domestic and foreign yields are listed in Table \ref{table4} and Table \ref{table5}. The parameters $(\kappa,\theta,\xi,\etabar)$ are shown in Table \ref{table3}, which are obtained by (roughly) calibrating a standard Heston model to the market option prices. In this case, $2\kappa\theta/\xi^2 = 0.169\ll 1$ and the Feller condition is strongly violated. 

\vspace{1em}
\begin{table}[H]
  \begin{center}
  \begin{tabular}{|c|cccccc|}
  \hline
    Parameter & $\kappa$ & $\theta$ & $\xi$ & $\etabar$ & $Z_0$ & $V_0$ \\ \hline
    Value & 0.8721 & 0.0276 & 0.5338 & -0.3566 & 0.2287 & 0.012 \\ \hline
  \end{tabular}
  \caption{The parameters of the OT-LSV model in the FX market data example.}
  \label{table3}
  \end{center}
\end{table}
\vspace{-1em}

For the numerical settings, the spatial computational domain is set to $D=[-0.6,1.0]\times[0,2]$ with $101$ points in each dimension. In order to improve the accuracy while still keeping a reasonable computation time, we employ a non-uniform mesh over $D$ and place more points around $(Z_0, V_0)$ \citep[see e.g.,][Section 2.2.]{foulon2010adi}. For the time interval $[0,5]$, we use 30 time steps with an equal step size between any two consecutive maturities, e.g., 30 time steps in $(0, 1/12]$ and 30 time steps in $(1/12, 1/6]$, and so on. Since there are 10 maturities (see Table \ref{table4}), we have 300 time steps for 5 years in total.

\begin{figure}[p]
  \begin{minipage}{0.49\textwidth}
    \centering
    \includegraphics[width=1.0\linewidth]{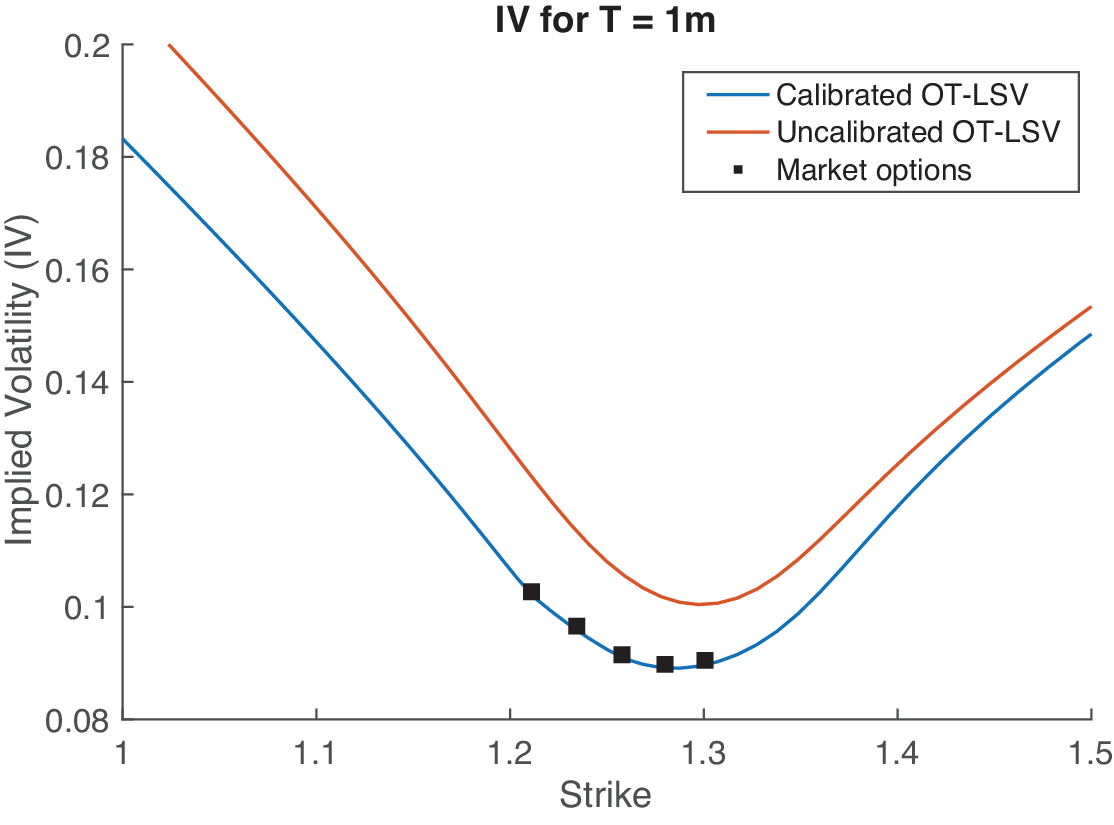}
  \end{minipage}\hfill
  \begin{minipage}{0.49\textwidth}
    \centering
    \includegraphics[width=1.0\linewidth]{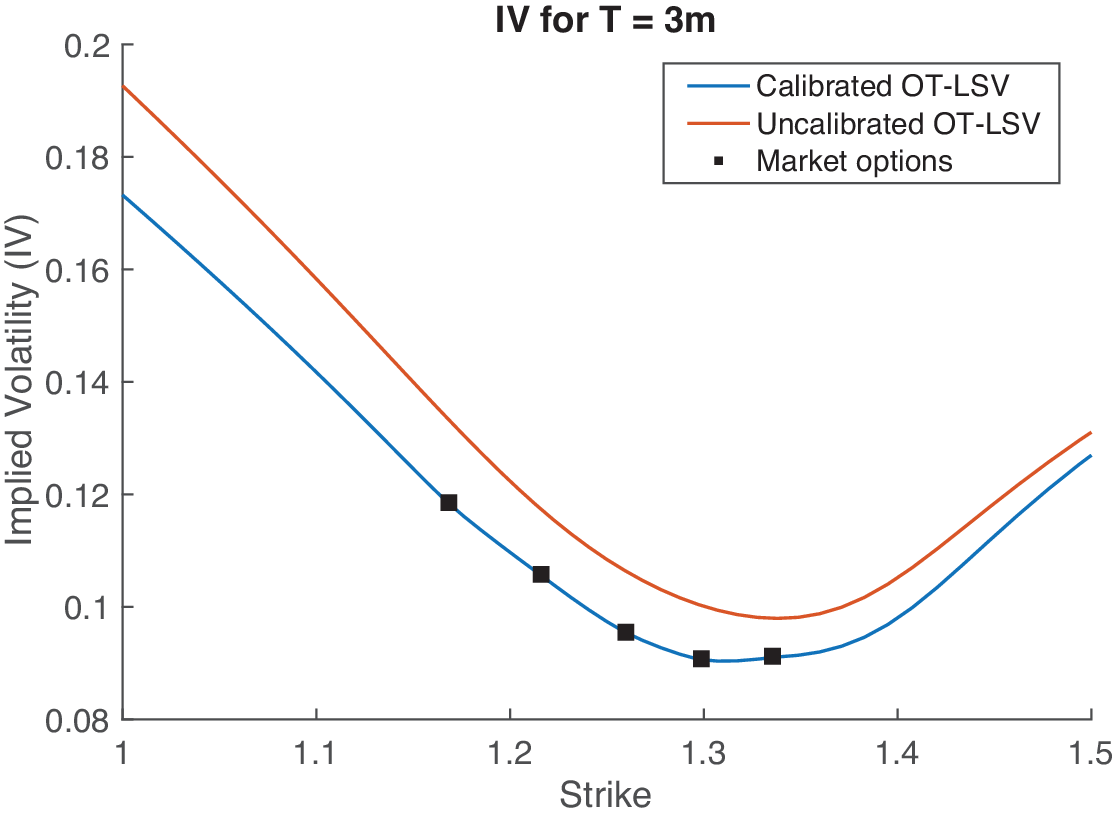}
  \end{minipage}
  \caption{The implied volatility (IV) skews generated by both the uncalibrated and the calibrated OT-LSV model for 1 month and 3 months maturities in the FX market data example.}
  \label{fig:short_IV}
\end{figure}

\begin{figure}[p]
  \begin{minipage}{0.49\textwidth}
    \centering
    \includegraphics[width=1.0\linewidth]{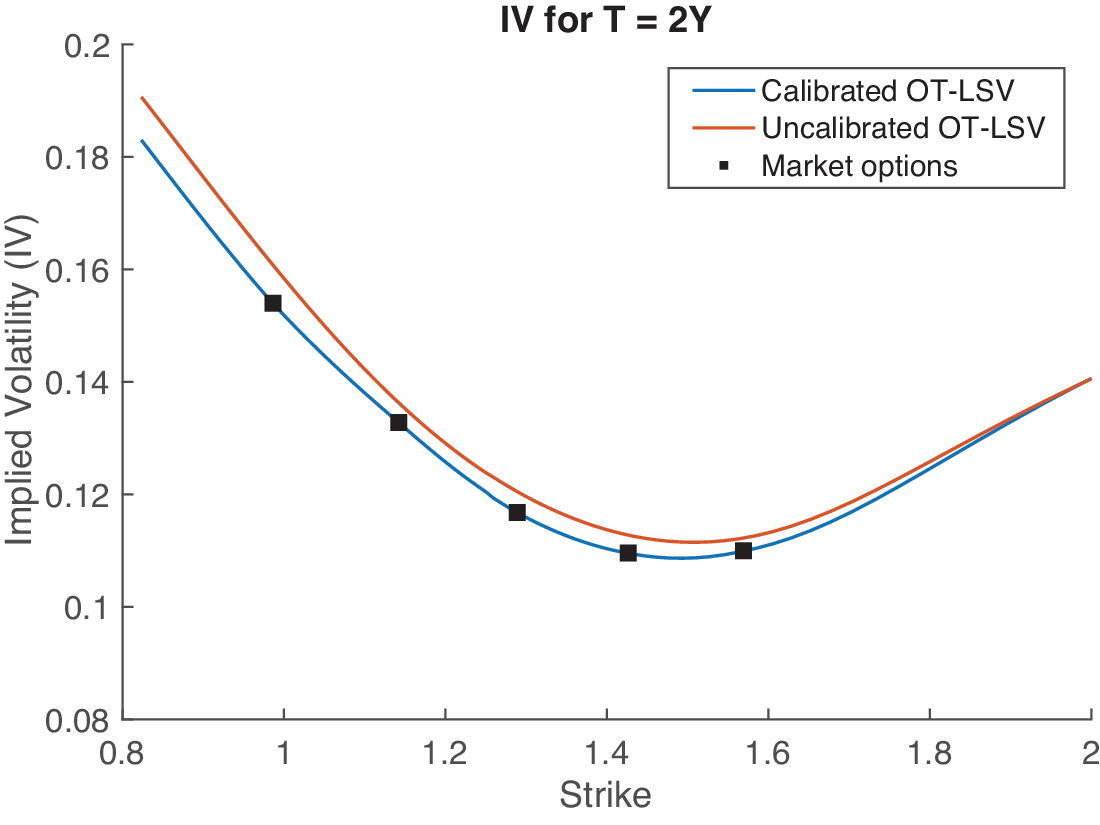}
  \end{minipage}\hfill
  \begin{minipage}{0.49\textwidth}
    \centering
    \includegraphics[width=1.0\linewidth]{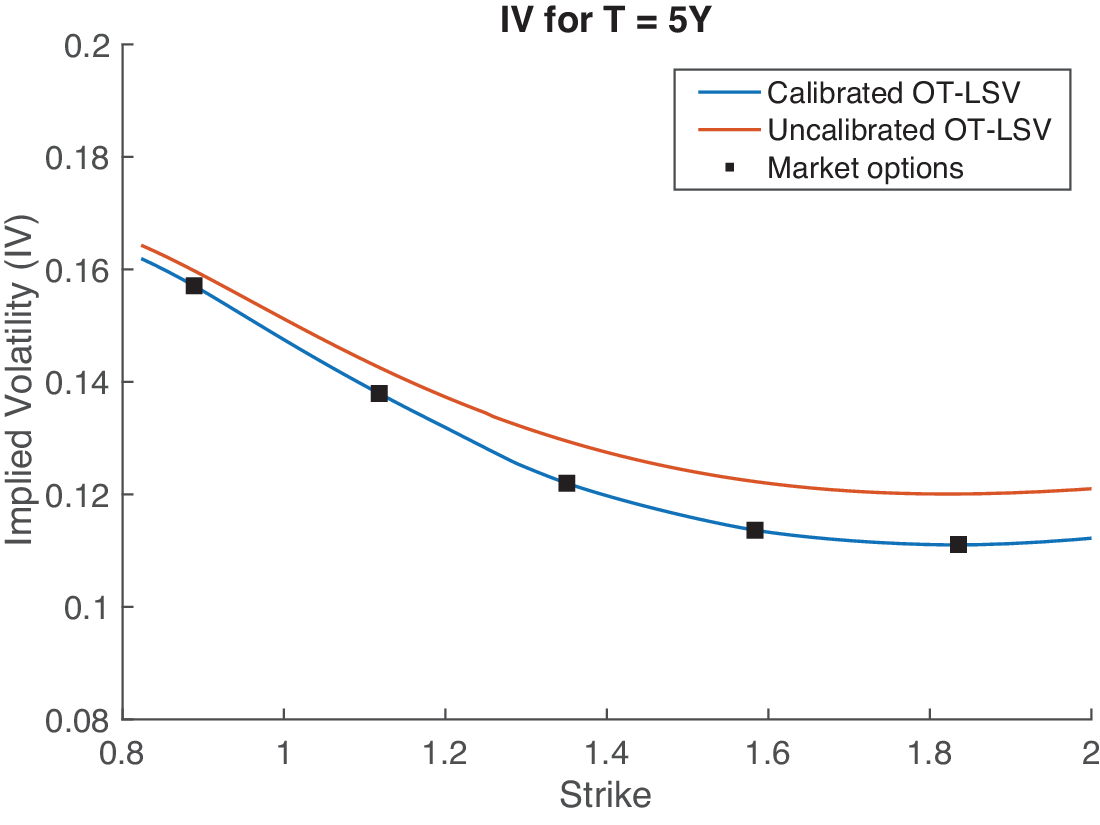}
  \end{minipage}
  \caption{The implied volatility (IV) skews generated by both the uncalibrated and the calibrated OT-LSV model for 2 years and 5 years maturities in the FX market data example.}
  \label{fig:long_IV}
\end{figure}

Setting a threshold of $\epsilon_1=6\times10^{-6}$, we obtain an exact calibration. The maximum difference between the model implied volatility and the market implied volatility is less than 1 basis point. Figure \ref{fig:short_IV} shows the implied volatility of the short-maturity options (1 month and 3 months) for the market data, the uncalibrated LSV model and the OT-calibrated LSV model. Figure \ref{fig:long_IV} shows the implied volatility of the long-maturity options (2 years and 5 years). 

\section*{Acknowledgements}

The Centre for Quantitative Finance and Investment Strategies has been supported by BNP Paribas. I. Guo has been partially supported by the Australian Research Council (Grant DP170101227). S. Wang has been supported by an Australian Government Research Training Program (RTP) Scholarship.

\clearpage
\appendix
\section{Appendix}

\subsection{Lemma \ref{lemma:A1}}

\begin{lemma}\label{lemma:A1}
Define $\Phi:C_b(\Lambda,\cX)\to\R\cup\{+\infty\}$ by
\eqn{
  \Phi(r,a,b) &= \left\{ \begin{array}{ll} 0 & \mbox{if } r+F^*(a,b)\leq 0, \\
			 +\infty & \mbox{otherwise.} \end{array} \right.
}
If we restrict the domain of its convex conjugate $\Phi^*:C_b(\Lambda,\cX)^*\to\R\cup\{+\infty\}$ to $\cM(\Lambda,\cX)$, then
\eqn{
  \Phi^*(\rho,\cA,\cB) &= \left\{ \begin{array}{ll} \displaystyle\int_\Lambda F\left(\frac{d\cA}{d\rho},\frac{d\cB}{d\rho}\right)\,d\rho & \mbox{if }\rho\in\cM_+(\Lambda) \mbox{ and } (\cA,\cB)\ll\rho, \\
			 +\infty & \mbox{otherwise.} \end{array} \right.
}
\end{lemma}
\begin{proof}
Let us identify the cases where $\Phi^*<+\infty$. For any $(\rho,\cA,\cB)\in C_b(\Lambda,\cX)^*$, using the definition of convex conjugate, we have
\eqn{
  \Phi^*(\rho,\cA,\cB) = \sup_{(r,a,b)\in C_b(\Lambda,\cX)}\{ \braket{(r,a,b)}{(\rho,\cA,\cB)} \;;\; r+F^*(a,b)\leq 0 \} .
}
If we restrict the domain of $\Phi^*$ to $\cM(\Lambda,\cX)\subset C_b(\Lambda,\cX)^*$, then
\eqn{
  \Phi^*(\rho,\cA,\cB) = \sup_{(r,a,b)\in C_b(\Lambda,\cX)}\bigg\{ \int_\Lambda r\,d\rho + a\cdot d\cA + b:d\cB \;;\; r+F^*(a,b)\leq 0 \bigg\}.
}
To show that one can restrict to $\rho\in\cM_+(\Lambda,\cX)$ if $\Phi^*<+\infty$, we assume that there exists a measurable set $E\subset\Lambda$ such that $\rho(E)<0$. By the fact that $C_b$ is dense in $L^1$, there exists a sequence of nonnegative functions $\zeta_n\in C_b(\Lambda)$ that converges to ${\mathds 1}_E\in L^1(d\rho_tdt)$. Let us construct a sequence $(r_n, a_n, b_n)=(-k\zeta_n,O^{d\times 1},O^{d\times d})\in C_b(\Lambda,\cX)$ where $k$ is an arbitrary positive constant and $O^{m\times n}$ denotes a null matrix of size $m\times n$. It is clear that the constraint $r+F^*(a, b)\leq 0$ is satisfied at $(r,a,b)=(r_n,a_n,b_n)$ as $F^*(O^{d\times 1}, O^{d\times d})\leq 0$. Then, by the dominated convergence theorem, we have
\eqn{
  \Phi^*(\rho,\cA,\cB) &\geq \lim_{n\to+\infty}\int_\Lambda r_n\,d\rho + a_n\cdot d\cA + b_n:d\cB \\
  &= \int_\Lambda \lim_{n\to+\infty} r_n\,d\rho\\
  &= -k\int_\Lambda \lim_{n\to+\infty} \zeta_n\,d\rho \\
  &= -k\rho(E).
}
If we send $k$ to infinity, the function $\Phi^*$ becomes unbounded.

To show that it is necessary to have $(\cA,\cB)\ll\rho$ if $\Phi^*<+\infty$, we assume that there exists a measurable set $E$ such that $(\cA,\cB)(E)\neq 0$ but $\rho(E)=0$. Again, by the fact that $C_b$ is dense in $L^1$, there exists a sequence of functions $\zeta_n\in C_b(\Lambda)$ such that $\zeta_n$ take values between $0$ and $1$ and the sequence converges to ${\mathds 1}_E\in L^1(d\rho_tdt)$. Such sequence can be found by taking convolution of ${\mathds 1}_E$ with a standard regularising kernel. Let us construct a sequence $(r_n,a_n,b_n) = (-F^*(k_1 I^{d\times 1}, k_2 I^{d\times d})\zeta_n, k_1\zeta_n I^{d\times 1}, k_2\zeta_n I^{d\times d})\in C_b(\Lambda,\cX)$ where $k_1,k_2$ are arbitrary constants and $I^{m\times n}$ denotes an all-ones matrix of size $m\times n$. By the convexity of $F^*$ and the fact that $F^*(O^{d\times 1}, O^{d\times d})\leq 0$, it is clear that the constraint $r+F^*(a, b)\leq 0$ is satisfied at $(r,a,b)=(r_n,a_n,b_n)$. Then, by the dominated convergence theorem, we have
\eqn{
  \Phi^*(\rho,\cA,\cB) &\geq \lim_{n\to+\infty}\int_\Lambda r_n\,d\rho + a_n\cdot d\cA + b_n:d\cB \\
  &= \int_\Lambda \lim_{n\to+\infty} r_n\,d\rho + \int_\Lambda \lim_{n\to+\infty} a_n\cdot d\cA + \int_\Lambda \lim_{n\to+\infty} b_n:d\cB \\
  &= -\int_\Lambda \lim_{n\to+\infty} F^*(k_1 I^{d\times 1},k_2I^{d\times d})\zeta_n\,d\rho + \int_\Lambda \lim_{n\to+\infty} k_1\zeta_n I^{d\times 1}\cdot d\cA + \int_\Lambda \lim_{n\to+\infty} k_2\zeta_n I^{d\times d}:d\cB\\
  &= k_1\sum_{i}(\cA(E))_i + k_2\sum_{i,j}(\cB(E))_{ij}
}
The function $\Phi^*$ goes to infinity if we send $k_1, k_2$ to $+\infty$ or $-\infty$, depending on the sign of $\sum_{i}(\cA(E))_i$ and $\sum_{i,j}(\cB(E))_{ij}$.

Now, since the integrand of the integral in $\Phi^*$ is linear in $(r,a,b)$, if $\Phi^*$ is finite, the supremum must occur at the boundary. Thus, assuming that $\rho\in\cM_+(\Lambda,\cX)$ and $(\cA,\cB)\ll\rho$, we have
\eqn{
  \Phi^*(\rho,\cA,\cB) &= \sup_{r+F^*(a,b)= 0}\int_\Lambda \left( r + a\cdot\frac{d\cA}{d\rho} + b:\frac{d\cB}{d\rho} \right) \,d\rho \\
     &= \sup_{(a,b)}\int_\Lambda \left( a\cdot\frac{d\cA}{d\rho} + b:\frac{d\cB}{d\rho} - F^*(a,b) \right) \,d\rho \\
     &\leq \int_\Lambda \sup_{(a,b)}\left( a\cdot\frac{d\cA}{d\rho} + b:\frac{d\cB}{d\rho} - F^*(a,b) \right) \,d\rho \\
     &= \int_\Lambda F\left(\frac{d\cA}{d\rho},\frac{d\cB}{d\rho}\right)\,d\rho.
}
The last equality holds since the convex and lower semi-continuous function $F$ coincides with its biconjugate $F^{**}$ according to the Fenchel--Moreau theorem \citep[see e.g.,][Theorem 1.11]{brezis2011}. 

Conversely, by the density of $C_b$ in $L^1$, let us choose a sequence of functions $(a_n, b_n)\in C_b(\Lambda, \R^d\times\bS^d)$ converging to $\nabla F(\frac{d\cA}{d\rho},\frac{d\cB}{d\rho})=\arg\sup_{(a,b)}\left( a\cdot\frac{d\cA}{d\rho} + b:\frac{d\cB}{d\rho} - F^*(a,b) \right)$ in $L^1(d\rho_tdt, \R^d\times\bS^d)$. Applying the dominated convergence theorem, we have
\eqn{
  \Phi^*(\rho,\cA,\cB) &= \sup_{(a,b)}\int_\Lambda \left( a\cdot\frac{d\cA}{d\rho} + b:\frac{d\cB}{d\rho} - F^*(a,b) \right) \,d\rho \\
  &\geq \lim_{n\to+\infty}\int_\Lambda \left( a_n\cdot\frac{d\cA}{d\rho} + b_n:\frac{d\cB}{d\rho} - F^*(a_n,b_n) \right) \,d\rho \\
  &= \int_\Lambda \lim_{n\to+\infty} \left( a_n\cdot\frac{d\cA}{d\rho} + b_n:\frac{d\cB}{d\rho} - F^*(a_n,b_n) \right) \,d\rho \\
  &= \int_\Lambda \sup_{(a,b)}\left( a\cdot\frac{d\cA}{d\rho} + b:\frac{d\cB}{d\rho} - F^*(a,b) \right) \,d\rho \\
  &= \int_\Lambda F\left(\frac{d\cA}{d\rho},\frac{d\cB}{d\rho}\right)\,d\rho.
}
The proof is completed.
\end{proof}

\clearpage
\subsection{Lemma \ref{lemma:A2}}

In this section, we prove that the duality between spaces $C_b$ and $\cM$ can be extended to the non-compact space $[0,T]\times\R^d$ in this particular case. A similar argument for the Kantorovich duality of the classical optimal transport was made in \citet[Appendix 1.3]{villani2003book}.

\begin{lemma}\label{lemma:A2}
  Denote by $K^o$ the set of $(r,a,b)$ in $C_b(\Lambda,\cX)$ that can be represented by some $(\phi,\lambda)$ in $BV([0,T],C_b^2(\R^d))\times\R^m$ with $\phi(T,\cdot)=0$ (see the proof of Theorem 3.5 for the definition of `represented').  Let $\Phi^*:C_b(\Lambda,\cX)^* \to \R\cup\{+\infty\}$ and $\Psi^*:C_b(\Lambda,\cX)^* \to \R\cup\{+\infty\}$ be defined by
  \eqn{
    \Phi^*(\rho,\cA,\cB) =& \sup_{(r,a,b)\in C_b(\Lambda,\cX)} \{ \braket{(r,a,b)}{(\rho,\cA,\cB)} \;;\; r+F^*(a,b)\leq 0 \}, \\
    \Psi^*(\rho,\cA,\cB) =& \sup_{(r,a,b)\in K^o} \left\{ \braket{(r,a,b)}{(\rho,\cA,\cB)} - \int_{\R^d} \phi(0,x)\,d\mu_0 + \sum_{i=1}^m \lambda_i c_i \right\}.
  }
  Then, 
  \eq{\label{eq:A2}
    \inf_{(\rho,\cA,\cB)\in C_b(\Lambda,\cX)^*} (\Phi^*+\Psi^*)(\rho,\cA,\cB) = \inf_{(\rho,\cA,\cB)\in \cM(\Lambda,\cX)} (\Phi^*+\Psi^*)(\rho,\cA,\cB).
  }
\end{lemma}
\begin{proof}
  Let $C_0(\Lambda,\cX)$ be the space of continuous functions on $\Lambda$ valued in $\cX$ that vanish at infinity. We decompose $(\rho,\cA,\cB) = (\trho,\tcA,\tcB) + (\delta\rho,\delta\cA,\delta\cB)$ such that $(\trho,\tcA,\tcB)\in\cM(\Lambda,\cX)$ and $\braket{(\phi_\rho,\phi_\cA,\phi_\cB)}{(\delta\rho,\delta\cA,\delta\cB)}=0$ for any $(\phi_\rho,\phi_\cA,\phi_\cB)\in C_0(\Lambda,\cX)$ (The reader can refer to \citet[Appendix 1.3]{villani2003book} for the existence of such a decomposition.). Since, $\cM(\Lambda,\cX)$ is a subset of $C_b(\Lambda,\cX)^*$, it follows that
\eqn{
  \inf_{(\rho,\cA,\cB)\in C_b(\Lambda,\cX)^*} (\Phi^*+\Psi^*)(\rho,\cA,\cB) \leq \inf_{(\rho,\cA,\cB)\in \cM(\Lambda,\cX)} (\Phi^*+\Psi^*)(\rho,\cA,\cB).
}

Next, we show that the converse of the above inequality is also valid. If $\Phi^*\equiv+\infty$ or $\Psi^*\equiv+\infty$, then the proof is trivial. Thus, we assume that $\Phi^*$ and $\Psi^*$ take finite values at some $(\rho,\cA,\cB)\in C_b(\Lambda,\cX)^*$. For $\Phi^*$, since $C_0(\Lambda,\cX)\subseteq C_b(\Lambda,\cX)$, we have
\eq{
  \Phi^*(\rho,\cA,\cB) &= \sup_{(r,a,b)\in C_b(\Lambda,\cX)} \{ \braket{(r,a,b)}{(\rho,\cA,\cB)} \;;\; r+F^*(a,b)\leq 0 \} \nonumber\\
    &\geq \sup_{(r,a,b)\in C_0(\Lambda,\cX)}\{ \braket{(r,a,b)}{(\rho,\cA,\cB)}  \;;\; r+F^*(a,b)\leq 0 \} \label{eq:lemma_a2_phi}\\
    &= \sup_{(r,a,b)\in C_0(\Lambda,\cX)}\bigg\{ \int_\Lambda r\,d\trho + a\cdot d\tcA + b:d\tcB \;;\; r+F^*(a,b)\leq 0 \bigg\}. \nonumber
}
Let $\chi_n\in C_0(\Lambda)$ be a sequence of cutoff functions with $0\leq \chi_n\leq 1$ on $\Lambda$ and $\chi_n\to 1$ as $n\to\infty$. The existence of the sequence $(\chi_n)$ follows from the Urysohn's lemma \citep[Lemma 2.12]{rudin1987real}. Let us construct a sequence $(r_n,a_n,b_n)=(-F^*(a, b)\chi_n, a\chi_n, b\chi_n)\in C_0(\Lambda,\cX)$ for some $(a,b)\in C_b(\Lambda,\R^d\times\bS^d)$, then $(r_n,a_n,b_n)\to(-F^*(a, b), a, b)\in C_b(\Lambda,\cX)$ as $n\to\infty$. The finiteness of $F^*(a,b)$ is guaranteed by the coercivity of $F$. By the convexity of $F^*$ and the fact that $F^*(O^{d\times 1}, O^{d\times d})\leq 0$ where $O^{m\times n}$ denotes a null matrix of size $m\times n$, it is clear that $(r_n,a_n,b_n)$ satisfies $r_n+F^*(a_n,b_n)\leq 0$. Since the supremum in the last line of \eqref{eq:lemma_a2_phi} is taken over all $(r,a,b)\in C_0(\Lambda, \cX)$, we have
\eqn{
  \Phi^*(\rho,\cA,\cB) &\geq \sup_{(a,b)\in C_b(\Lambda,\R^d\times\bS^d)}\lim_{n\to\infty} \bigg\{ \int_\Lambda r_n\,d\trho + a_n\cdot d\tcA + b_n:d\tcB \bigg\} \\
  &= \sup_{(a,b)\in C_b(\Lambda,\R^d\times\bS^d)} \bigg\{ \int_\Lambda -F^*(a,b)\,d\trho + a\cdot d\tcA + b:d\tcB \bigg\} \\
  &= \sup_{(r, a,b)\in C_b(\Lambda,\cX)} \bigg\{ \int_\Lambda r\,d\trho + a\cdot d\tcA + b:d\tcB \;;\; r+F^*(a,b)\leq 0 \bigg\} \\
  &= \Phi^*(\trho, \tcA, \tcB).
}
The first equality above is justified by the dominated convergence theorem. The second equality above holds because if $\Phi^*$ is finite, then the supremum must occur at the boundary. 

For $\Psi^*$, if we restrict its domain to $(\trho,\tcA,\tcB)\in\cM(\Lambda,\cX)$, then $\Psi^*=0$ if $(\trho,\tcA,\tcB)$ satisfies (\ref{eq:weak_cons_1}) and (\ref{eq:weak_cons_2}) or $\Psi^*=+\infty$ otherwise. Recall that in $K^o$, $r=-\dt\phi-\sum_{i=1}^m\lambda_i G_i \delta_i$, $a=-\Dx\phi$ and $b=-\demi\Dxx\phi$. Whenever $\Psi^*$ is finite, by (\ref{eq:weak_cons_1}) and (\ref{eq:weak_cons_2}), we have
\eq{\label{eq:A22}
  \int_\Lambda  r\,d\trho + a\cdot d\tcA + b:d\tcB -\int_{\R^d}\phi(0,x)\,d\mu_0 + \sum_{i=1}^m \lambda_i c_i  = 0 \qquad \forall (r,a,b)\in K^o.
}
The equation (\ref{eq:A22}) holds in particular for $(r,a,b)$ in the subset $K^o\cap C_0(\Lambda,\cX)$. Also, since $K^o\cap C_0(\Lambda,\cX) \subseteq K^o$, we have
\eqn{
  \Psi^*(\rho,\cA,\cB) &= \sup_{(r,a,b)\in K^o}\left\{ \braket{(r,a,b)}{(\rho,\cA,\cB)} - \int_{\R^d} \phi(0,x)\,d\mu_0 + \sum_{i=1}^m \lambda_i c_i \right\}\\
  &\geq \sup_{(r,a,b)\in K^o\cap C_0(\Lambda,\cX)}\left\{ \braket{(r,a,b)}{(\rho,\cA,\cB)} - \int_{\R^d} \phi(0,x)\,d\mu_0 + \sum_{i=1}^m \lambda_i c_i \right\}\\
  &= \sup_{(r,a,b)\in K^o\cap C_0(\Lambda,\cX)}\left\{ \int_\Lambda  r\,d\trho + a\cdot d\tcA + b:d\tcB - \int_{\R^d} \phi(0,x)\,d\mu_0 + \sum_{i=1}^m \lambda_i c_i \right\}\\
  &= \sup_{(r,a,b)\in K^o}\left\{ \int_\Lambda  r\,d\trho + a\cdot d\tcA + b:d\tcB - \int_{\R^d} \phi(0,x)\,d\mu_0 + \sum_{i=1}^m \lambda_i c_i \right\}\\
  &= \Psi^*(\trho,\tcA,\tcB).
}
Therefore,
\eqn{
  \inf_{(\rho,\cA,\cB)\in C_b(\Lambda,\cX)^*} (\Phi^*+\Psi^*)(\rho,\cA,\cB) \geq \inf_{(\rho,\cA,\cB)\in \cM(\Lambda,\cX)} (\Phi^*+\Psi^*)(\rho,\cA,\cB).
}
This completes the proof.
\end{proof}

\clearpage
\subsection{Algorithm 2}\label{sec:alg2}

\begin{algorithm}[H]
  \DontPrintSemicolon
  \SetNoFillComment
  \setcounter{AlgoLine}{0}
  \KwData{Market prices of European option}
  \KwResult{A calibrated OT-LSV model that matches all market prices}
  
  Set an initial $\lambda$\;
  \Do{$\norm{\nabla J(\lambda)}_\infty > \epsilon_1$}{
    \For{$k = N_T-1,\ldots,0$}{
      \tcc{Solving the HJB equation}
      \If{$t_{k+1}$ is equal to the maturity of any calibrating options}{
        $\phi_{t_{k+1}} \gets \phi_{t_{k+1}} + \sum_{i=1}^m\lambda_i G_i{\mathds 1}(t_{k+1}=\tau_i)$\;
      }
      \tcc{Policy iteration}
      Let $\phi_{t_k}^{new} = \phi_{t_{k+1}}$\;
      \Do{$\norm{\phi_{t_k}^{new} - \phi_{t_k}^{old}}_2 > \epsilon_2$}{
        $\phi_{t_k}^{old} \gets \phi_{t_k}^{new}$\;
        Approximate $\sigma_{t_k}^2$ by solving (\ref{eq:lsv_optimal_sigma}) with $\phi=\phi_{t_k}^{old}$\;
        Solve the HJB equation (\ref{eq:lsv_hjb}) by the ADI method at $t=t_k$, and set the solution to $\phi_{t_k}^{new}$\;
      }
      $\phi_{t_k} \gets \phi_{t_k}^{new}$\;
    }
    \tcc{Calculating model prices and gradient}
    Solve (\ref{eq:pricing_PDE_alt}) to calculate the model prices by the ADI method\;
    Calculate the gradient $\nabla J(\lambda)$ by (\ref{eq:lsv_gradient})\;
    Update $\lambda$ by the L-BFGS algorithm\;
  }
 \caption{LSV calibration with policy iteration}
  \label{alg2}
\end{algorithm}

\clearpage
\subsection{FX options data}\label{sec:fxdata}

\begin{table}[H]
\begin{center}
\begin{tabular}{|c|c|c|c|c|c|c|c|}
\hline
Maturity  & Option type  & Strike & Implied Vol & Maturity  & Option type  & Strike & Implied Vol     \\ \hline
    & Call & 1.3006 & 0.0905 &    & Call & 1.4563 & 0.1069 \\ 
    & Call & 1.2800 & 0.0898 &    & Call & 1.3627 & 0.1052 \\ 
  1m & Call & 1.2578 & 0.0915 & 1Y & Call & 1.2715 & 0.1118 \\ 
    & Put  & 1.2344 & 0.0966 &    & Put  & 1.1701 & 0.1278 \\ 
    & Put  & 1.2110 & 0.1027 &    & Put  & 1.0565 & 0.1491 \\ \hline
    & Call & 1.3191 & 0.0897 &    & Call & 1.5691 & 0.1100 \\ 
    & Call & 1.2901 & 0.0896 &    & Call & 1.4265 & 0.1096 \\ 
  2m & Call & 1.2588 & 0.0933 & 2Y & Call & 1.2889 & 0.1168 \\ 
    & Put  & 1.2243 & 0.1014 &    & Put  & 1.1421 & 0.1328 \\ 
    & Put  & 1.1882 & 0.1109 &    & Put  & 0.9863 & 0.1540 \\ \hline
    & Call & 1.3355 & 0.0912 &    & Call & 1.6683 & 0.1109 \\ 
    & Call & 1.2987 & 0.0908 &    & Call & 1.4860 & 0.1122 \\ 
  3m & Call & 1.2598 & 0.0955 & 3Y & Call & 1.3113 & 0.1200 \\ 
    & Put  & 1.2160 & 0.1058 &    & Put  & 1.1308 & 0.1352 \\ 
    & Put  & 1.1684 & 0.1185 &    & Put  & 0.9468 & 0.1547 \\ \hline
    & Call & 1.3775 & 0.0960 &    & Call & 1.7507 & 0.1104 \\ 
    & Call & 1.3213 & 0.0953 &    & Call & 1.5351 & 0.1127 \\ 
  6m & Call & 1.2633 & 0.1013 & 4Y & Call & 1.3306 & 0.1210 \\ 
    & Put  & 1.1973 & 0.1145 &    & Put  & 1.1226 & 0.1365 \\ 
    & Put  & 1.1236 & 0.1316 &    & Put  & 0.9152 & 0.1554 \\ \hline
    & Call & 1.4068 & 0.1013 &    & Call & 1.8355 & 0.1111 \\ 
    & Call & 1.3329 & 0.1005 &    & Call & 1.5835 & 0.1137 \\ 
  9m & Call & 1.2583 & 0.1068 & 5Y & Call & 1.3505 & 0.1220 \\ 
    & Put  & 1.1745 & 0.1215 &    & Put  & 1.1180 & 0.1379 \\ 
    & Put  & 1.0805 & 0.1407 &    & Put  & 0.8887 & 0.1571 \\ \hline
\end{tabular}
\caption{The EUR/USD option data as of 23 August 2012. The spot price $S_0 = 1.257$ USD per EUR. At each maturity, the options correspond to 10-delta calls, 25-delta calls, 50-delta calls, 25-delta puts and 10-delta puts}
\label{table4}
\end{center}
\end{table}

\begin{table}[H]
\begin{center}
\begin{tabular}{|c|cccccccccc|}
\hline
  Maturity & 1m & 2m & 3m & 6m & 9m & 1Y & 2Y & 3Y & 4Y & 5Y  \\ \hline
  Domestic yield & 0.41 & 0.51 & 0.66 & 0.95 & 1.19 & 1.16 & 0.60 & 0.72 & 0.72 & 0.72  \\ \hline
  Foreign yield & 0.04 & 0.11 & 0.23 & 0.47 & 1.62 & 0.64 & 0.03 & 0.03 & 0.03 & 0.03  \\ \hline
\end{tabular}
\caption{The domestic and foreign yields (in \%) as of 23 August 2012.}
\label{table5}
\end{center}
\end{table}

\clearpage
\bibliographystyle{apalike}
\bibliography{ref}

\end{document}